\documentclass[onecolumn,11pt]{article}
\usepackage[top=1in, bottom=1in, left=1in, right=1in]{geometry}
\usepackage{graphicx}
\usepackage{epstopdf}
\usepackage{color}
\usepackage{multirow}
\usepackage{subeqnarray}
\usepackage{setspace}
\usepackage[table]{xcolor}
\usepackage{palatino}
\usepackage{dcolumn}
\usepackage{bm}
\usepackage[utf8]{inputenc}
\usepackage[T1]{fontenc}
\usepackage{mathptmx}
\usepackage{amsmath,amsfonts,amscd,amssymb,amsthm}
\usepackage{commath}
\usepackage{wrapfig}

\usepackage{tcolorbox}
\tcbuselibrary{theorems}
\newtcbtheorem[]{mytheo}{Theorem}%
{colback=green!5,colframe=green!35!black,fonttitle=\bfseries}{th}

\usepackage{algorithm,algcompatible,algorithmicx}
\usepackage{algpseudocode}
\renewcommand{\Comment}[2][.4\linewidth]{%
	\leavevmode\hfill\makebox[#1][l]{$\triangleright$~#2}}
\algnewcommand\algorithmicinput{\textbf{Input:}}
\algnewcommand\INPUT{\item[\algorithmicinput]}
\algnewcommand\algorithmicoutput{\textbf{Output:}}
\algnewcommand\OUTPUT{\item[\algorithmicoutput]}
\algnewcommand\algorithmicoptional{\textbf{Optional:}}
\algnewcommand\OPTIONAL{\item[\algorithmicoptional]}
\usepackage{graphicx}
\usepackage{epstopdf}
\usepackage{overpic}
\usepackage{cancel}
\usepackage{rotating}
\usepackage{url}
\usepackage{caption}
\usepackage{subcaption}
\usepackage{xcolor}
\usepackage{rotating}
\usepackage{mathtools}

\DeclareMathOperator*{\argmin}{arg\,min}
\usepackage{setspace}
\usepackage[titletoc,title]{appendix}
\usepackage[numbers,sort&compress]{natbib}

\usepackage[bottom,flushmargin,hang,multiple,symbol]{footmisc}
\usepackage{lipsum}
\newcommand\blfootnote[1]{%
  \begingroup
  \renewcommand\thefootnote{}\footnote{#1}%
  \addtocounter{footnote}{-1}%
  \endgroup
}

\definecolor{header1}{cmyk}{0,0,0,1}

\title{\vspace{-.5in}\huge{\textbf{Promoting global stability in data-driven models\\ of quadratic nonlinear dynamics}}\vspace{-0.15in}}

\author{
\normalsize{Alan A. Kaptanoglu$^{*}$, Jared L. Callaham, Christopher J. Hansen,}  \normalsize{Aleksandr Aravkin, Steven L. Brunton}\\
\footnotesize{University of Washington, Seattle, WA 98195, United States}\vspace{-1in}
}
\date{}

 \normalsize
 
\begin{document} 
 \maketitle
 \blfootnote{$^*$ Corresponding author (akaptano@uw.edu).}
 \vspace{-0.15in}

 \begin{abstract}
 Modeling realistic fluid and plasma flows is computationally intensive, motivating the use of reduced-order models for a variety of scientific and engineering tasks. 
 However, it is challenging to characterize, much less guarantee, the global stability (i.e., long-time boundedness) of these models. 
 The seminal work of Schlegel and Noack~\cite{Schlegel2015jfm} provided a theorem outlining necessary and sufficient conditions to ensure global stability in systems with energy-preserving, quadratic nonlinearities, with the goal of evaluating the stability of projection-based models. 
 In this work, we incorporate this theorem into modern data-driven models obtained via machine learning. 
 First, we propose that this theorem should be a standard diagnostic for the stability of projection-based and data-driven models, examining the conditions under which it holds.
 Second, we illustrate how to modify the objective function in machine learning algorithms to promote globally stable models, with implications for the modeling of fluid and plasma flows. 
 Specifically, we introduce a modified ``trapping SINDy'' algorithm based on the sparse identification of nonlinear dynamics (SINDy) method.
 This method enables the identification of models that, by construction, only produce bounded trajectories. 
 The effectiveness and accuracy of this approach are demonstrated on a broad set of examples of varying model complexity and physical origin, including the vortex shedding in the wake of a circular cylinder. \\
 \noindent\textbf{Keywords: fluid mechanics, reduced-order modeling, Galerkin projection, machine learning, global stability, nonlinear systems, magnetohydrodynamics, data-driven models, SINDy } 
 \end{abstract}
\vspace{-0.2in}
\section{Introduction\label{sec:intro}}
\begin{figure}[t]
\vspace{-.05in}
  \begin{center}
    \begin{overpic}[width=0.99\textwidth]{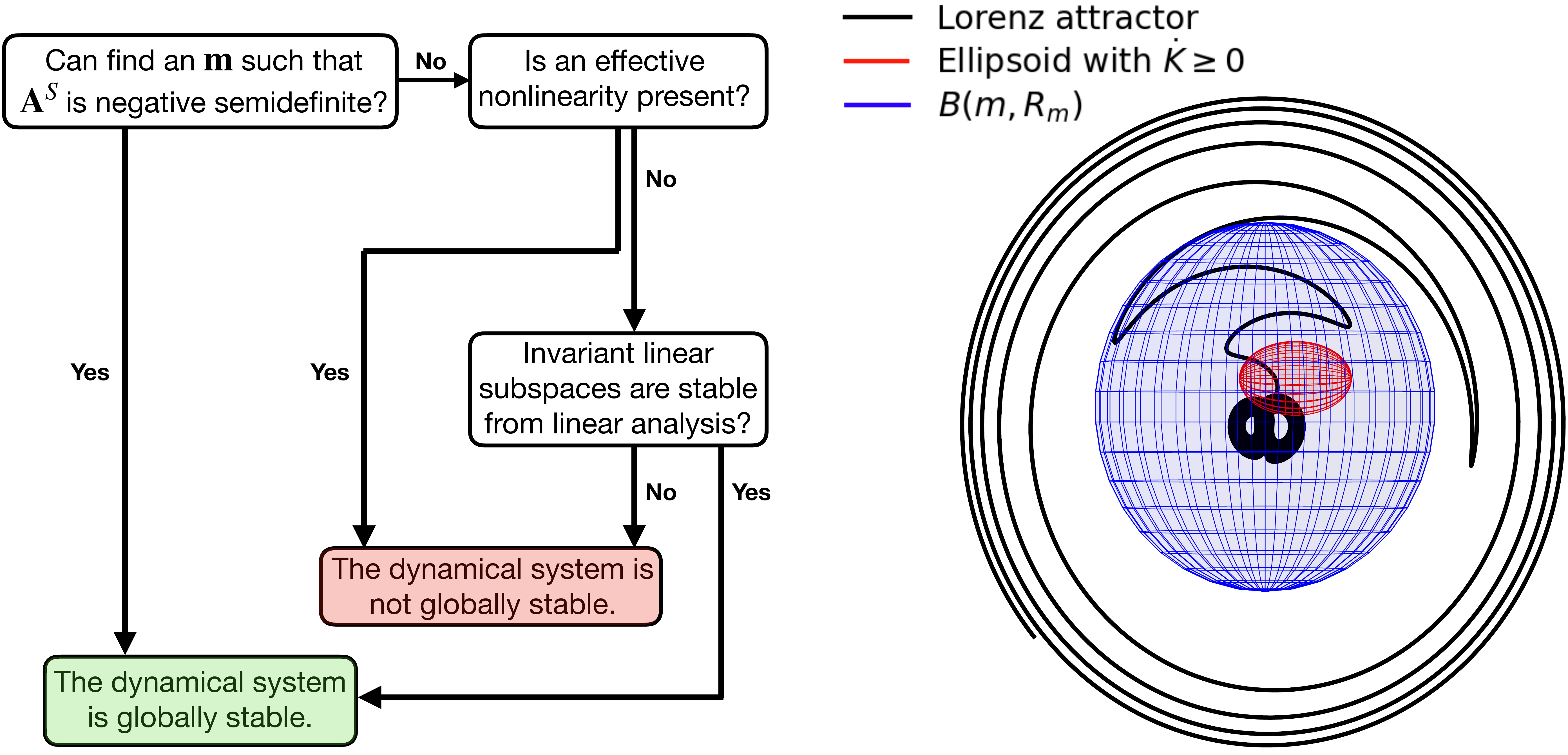}
    \end{overpic}
  \end{center}
  \vspace{-.22in}
  \caption{Left: Decision diagram to determine global stability, modified from Schlegel and Noack~\cite{Schlegel2015jfm} and described in Section~\ref{sec:trapping_theorem}. 
  Right: Illustration of a trapping region (blue sphere) for the Lorenz system; all outside trajectories monotonically approach this region, and after entering, remain inside. 
  Trajectories inside the red ellipsoid experience positive energy growth, in this case precluding convergence to a fixed point.}
  \label{fig:overview}
  \vspace{-0.1in}
\end{figure}

Modeling the full spatio-temporal evolution of natural processes is often computationally expensive, motivating the use of reduced-order models (ROMs) that capture only the dominant behaviors of a system~\cite{noack2003hierarchy,Noack2011book,carlberg2011efficient,carlberg2013gnat,benner2015survey,Rowley2017arfm}. 
Projection-based model reduction is a common approach for generating such models; a high-dimensional system, such as a spatially discretized set of partial differential equations (PDEs), is projected onto a low-dimensional basis of modes~\cite{Taira2017aiaa,taira2020modal}.  
This projection leads to a computationally efficient system of ordinary differential equations (ODEs) that describes how the mode amplitudes evolve in time~\cite{holmes2012turbulence}. 
However, these models often suffer from stability issues, causing solutions to diverge in finite-time.  
To address this issue, Schlegel and Noack~\cite{Schlegel2015jfm} developed a ``trapping theorem'' with necessary and sufficient conditions for long-term model stability for systems that exhibit quadratic, energy-preserving nonlinearities. 
Quadratic nonlinearity is pervasive in nature, with common examples including convection in the Navier-Stokes equations and the Lorentz force in magnetohydrodynamics (MHD). 
The trapping theorem provides conditions for the existence of a global trapping region, towards which every system trajectory asymptotically and monotonically converges; once a trajectory enters this region, it remains inside for all time, guaranteeing that all trajectories are bounded. 
These types of guarantees are ideal for the application of real-time flow-control strategies. 
An example trapping region is illustrated by the blue sphere in Fig.~\ref{fig:overview} for the Lorenz system.
For convenience, we will use the terms ``global stability'', ``long-term boundedness'', and ``monotonically trapping region'' interchangeably, although systems exhibiting trapping regions are a strict subset of globally stable systems (see Fig.~1 of Schlegel and Noack~\cite{Schlegel2015jfm} for a useful organizational diagram of these various notions of stability). 
In this work, we adapt the trapping theorem from projection-based modeling to promote global stability in data-driven machine learning models. 

Increasingly, reduced-order models of complex systems, such as fluids and plasmas, are discovered from data with modern machine learning algorithms~\cite{schmid_dynamic_2010,Rowley2009jfm,mezic_analysis_2013,Kaiser2014jfm,Brunton2016pnas,klus2018data,Rudy2017sciadv,Dam2017pf,loiseau2018constrained,Towne2018jfm,deng2020low,raissi2018hidden,pathak2018model,bar2019learning,Duraisamy2019arfm,rackauckas2020universal,alves2020data,pan2020physics,qian2020lift,kaptanoglu2020physics,li2020fourier,lee2020model,Herrmann2020arxiv,sanchez2020learning,kochkov2021machine,kaptanoglu2021structure}, rather than classical projection-based methods that are intrusive and require intricate knowledge of the governing equations. 
These data-driven approaches for modeling fluid dynamics~\cite{brenner2019perspective,brunton2020machine} range from generalized regression techniques~\cite{schmid_dynamic_2010,Brunton2016pnas,loiseau2018constrained} to deep learning~\cite{Duraisamy2019arfm,lee2020model,li2020fourier,sanchez2020learning,kochkov2021machine,raissi2020science}. 
It is often possible to improve the stability and performance of data-driven models by incorporating partially known physics, such as conservation laws and symmetries~\cite{loiseau2018constrained,kaptanoglu2020physics,kaptanoglu2021structure}, or known physical structure~\cite{cranmer2020lagrangian}. 
Thus, incorporating physics into machine learning and developing hybrid data-driven and operator-based approaches are rapidly growing fields of research~\cite{Majda2012nonlinearity,ballarin2015supremizer,peherstorfer2016data,loiseau2018constrained,yang2020physics,Raissi2019jcp,mohebujjaman2019physically,Noe2019science,lee2019deep,cranmer2020lagrangian}.
Physics can be incorporated into machine learning algorithms through model structure, by augmenting training data with known symmetries, by adding constraints to the optimization, or by adding custom loss functions~\cite{brunton2020machine}.  
However, even physics-informed data-driven models often lack global stability guarantees, and the ability of these methods to find long-term bounded models depreciates as the state dimension increases. 

In this work, we use the Schlegel and Noack~\cite{Schlegel2015jfm} trapping theorem to diagnose and promote global stability of data-driven models with quadratic nonlinearities. 
Even though their theorem was developed in the context of projection-based ROMs, we emphasize that it can be applied directly to analyze data-driven model stability \emph{post hoc}, and we examine conditions under which it holds.  
Next, we describe how to use this theorem to promote global stability in machine learned models by modifying the optimization loss function.   
We illustrate this approach on the sparse identification of nonlinear dynamics (SINDy) algorithm~\cite{Brunton2016pnas,Rudy2017sciadv} by implementing a custom optimization loss term that promotes models that are globally stable by construction.   
A constrained version of the SINDy optimization was previously developed to enforce energy-preserving quadratic structure in incompressible fluids~\cite{loiseau2018constrained} and it has since been extended for arbitrary state size and global conservation laws in magnetohydrodynamic systems~\cite{kaptanoglu2020physics,kaptanoglu2021structure}.
These constrained SINDy variants generally produce more stable models, and reflect a broader trend that stability issues in system identification can often be improved by building physical constraints into system identification methods~\cite{loiseau2018constrained,champion2020unified}. 
Our ``trapping SINDy'' algorithm generalizes previous stabilized or constrained reduced-order models for fluids by considering global rather than local stability, allowing for both transients and long-time attracting sets. 
Promoting global stability also improves robustness to noise over unconstrained or constrained SINDy. 
Recent work by Erichson et al.~\cite{erichson2019physics} promotes a more restrictive Lyapunov stable origin in fluid flows by adding a similar loss term to the optimization problem. 
Additionally, much of the literature has focused on the long-time energy properties of a dynamic attractor~\cite{balajewicz2013low} by either prescribing that the system be \textit{fully} energy-preserving (or Hamiltonian)~\cite{balajewicz2013lyapunov,carlberg2015preserving,peng2016symplectic,afkham2017structure, bhat2019learning,chu2020discovering} or applying real-time control~\cite{lasagna2016sum}. 
Mohebujjaman et al.~\cite{mohebujjaman2019physically} also used a simple version of the trapping theorem in order to constrain a hybrid projection-based and data-driven method.
The present work builds on these studies, providing a framework for addressing the long-standing challenge of promoting global stability in data-driven models.

The remainder of this paper is organized as follows: in Section~\ref{sec:roms}, we introduce the general class of systems with energy-preserving quadratic nonlinearities, investigate the circumstances under which the trapping theorem holds, and indicate connections with other stability descriptions in fluid mechanics. 
In Section~\ref{sec:methodology}, we define our ``trapping SINDy'' algorithm. 
Our trapping SINDy implementation is open-source and available through the PySINDy software package~\cite{silva2020pysindy}. 
This is a rather technical section on nonconvex optimization; the reader may skip this section and proceed to the results if the algorithmic details are not of interest. 
In Section~\ref{sec:results}, we demonstrate the effectiveness of this new system identification technique on a wide range of examples. Abridged versions of all of the results have been incorporated into a single PySINDy example notebook and can be reproduced in a few minutes on a laptop. In Section~\ref{sec:conclusion}, we conclude with suggestions for future work. 
Similar trapping theorems are promising for data-driven models in fields such as neuroscience, epidemiology, and population dynamics.

\section{Reduced-order modeling and the trapping theorem \label{sec:roms}}

Before describing how we incorporate the trapping theorem of Schlegel and Noack~\cite{Schlegel2015jfm} into data-driven models, here we briefly describe the family of projection-based ROMs for which the trapping theorem was introduced, and investigate the circumstances under which this theorem holds.
It is helpful to first motivate this work by reviewing the many scenarios under which energy-preserving quadratic nonlinearities can arise. In fluid dynamics, the quadratic nonlinearity often represents the convective derivative $(\bm{u}\cdot\nabla)\bm{u}$ in the Navier-Stokes equations. This quadratic nonlinearity is energy-preserving for a large number of boundary conditions. Examples include no-slip conditions, periodic boundary conditions~\cite{mccomb1990physics,holmes2012turbulence}, mixed no-slip and periodic boundary conditions~\cite{rummler1998direct}, and open flows in which the velocity magnitude decreases faster than the relevant surface integrals expand (e.g., two-dimensional rigid body wake flows and three-dimensional round
jets)~\cite{schlichting2016boundary}. 
In magnetohydrodynamics, there are additional quadratic nonlinearities through $\nabla\times(\bm{u}\times\bm{B})$ and $\bm{J}\times{\bm{B}}$, which are also energy-preserving with common experimental boundary conditions such as a conducting wall~\cite{freidberg2014ideal}, or a balance between dissipation and actuation in a steady-state plasma device~\cite{kaptanoglu2020two,kaptanoglu2021structure}. Notably, dissipationless Hall-MHD has four invariants; energy, cross-helicity, magnetic helicity, and generalized helicity~\cite{galtier2016introduction}, providing a wealth of potential model constraints for Hall-MHD ROMs. Here $\bm{u}$ is the fluid velocity, $\bm{J}$ is the electromagnetic current, and $\bm{B}$ is the magnetic field. 

\subsection{Projection-based ROMs}\label{Sec:ProjROMS}

In modern scientific computing, a set of governing partial differential equations is typically discretized into a high-dimensional system of coupled ordinary differential equations. 
In this work we will explicitly consider dynamics with linear plus quadratic structure, as are found in many fluid and plasma systems:
\begin{align}
    \dot{\bm{u}} = \bm{L}^0\bm{u} + \bm{Q}^0(\bm{u}).
\end{align}
Here we assume that the PDE has already been discretized for numerical computation, resulting in a coupled system of $n$ differential equations. 
The state of the system $\bm{u}(\bm{x},t)\in\mathbb{R}^n$ is a high-dimensional vector that represents the fluid velocity or other set of spatio-temporal fields, for example sampled on a high-resolution spatial grid. 
Thus, $\bm{L}^0$ and $\bm{Q}^0$ are high-dimensional operators used to perform the numerical simulation. The zero subscript distinguishes these operators from the Galerkin coefficients defined below in Eq.~\eqref{eq:Galerkin_model}.  

The goal of a projection-based ROM is to transform this high-dimensional system into a lower-dimensional system of size $r\ll n$ that captures the essential dynamics.  
One way to reduce the set of governing equations to a set of ordinary differential equations is by decomposition into a desired low-dimensional basis $\{\bm{\varphi}_i(\bm x)\}$ in a process commonly referred to as Galerkin expansion: 
\begin{align}
\label{eq:galerkin_expansion}
    \bm{u}(\bm{x}, t) = \overline{\bm{u}}(\bm{x}) + \sum_{i=1}^r a_i(t) \bm{\varphi}_i(\bm{x}).
\end{align}
Here, $\overline{\bm{u}}(\bm{x})$ is the mean field, $\bm{\varphi}_i(\bm x)$ are spatial modes, and $a_i(t)$ describe how the amplitude of these modes vary in time. 
The proper orthogonal decomposition (POD)~\cite{holmes2012turbulence,brunton2019data} is frequently used to obtain the basis, since the modes $\bm{\varphi}_i(\bm x)$ are orthogonal. 
Many other modal expansions and bases have been introduced for reduced-order fluid~\cite{Taira2017aiaa, taira2020modal} and plasma models~\cite{vskvara2020detection,ferreira2020deep,nayak2020dynamic,kaptanoglu2020}, including balanced POD~\cite{willcox2002balanced,rowley2005model}, spectral POD~\cite{Towne2018jfm}, dynamic mode decomposition (DMD)~\cite{schmid_dynamic_2010,Rowley2009jfm,Kutz2016book}, the Koopman decomposition~\cite{koopman_hamiltonian_1931,mezic_analysis_2013,pan2020sparsity}, resolvent analysis~\cite{mckeon2010critical,luhar2014opposition}, and autoencoders~\cite{lusch2018deep,champion2019data, lee2020model}. 
The governing equations are then Galerkin projected onto the basis $\{\bm{\varphi}_i(\bm x)\}$ by substituting Eq.~\eqref{eq:galerkin_expansion} into the PDE and using inner products to remove the spatial dependence.  
Orthogonal projection onto POD modes is the simplest and most common procedure, resulting in \emph{POD-Galerkin} models, although Petrov-Galerkin projection~\cite{carlberg2011efficient,carlberg2013gnat} has been shown to improve model performance in some cases.  
If the governing equations for $\bm{u}(\bm{x},t)$ are at most quadratic in nonlinearity, Galerkin projection produces the following system of ODEs for the set of temporal functions $ a_i(t)$,
\begin{align}
\label{eq:Galerkin_model}
\dot{a}_i(t) &= E_i+ \sum_{j=1}^rL_{ij}a_j + \sum_{j,k=1}^r Q_{ijk}a_ja_k.
\end{align}
$E_i$, $L_{ij}$, and $Q_{ijk}$ are tensors of static coefficients, obtained from spatial inner products between the $\bm{\varphi}_i(\bm x)$ and the operators $\bm{L}^0$ and $\bm{Q}^0$, that define the model dynamics. The class of systems we consider are those with energy-preserving nonlinearity, for which
\begin{align}
\label{eq:energy_preserving_nonlinearity_full}
    \sum_{i,j,k=1}^rQ_{ijk}a_ia_ja_k = 0, 
\end{align}
or equivalently, for all $i,j,k \in \{1,...,r\}$,
\begin{align}
\label{eq:energy_preserving_nonlinearity}
Q_{ijk} + Q_{jik} + Q_{kji}= 0.
\end{align} 
$Q_{ijk}$ is symmetric in swapping $j$ and $k$ without loss of generality.

\subsection{Schlegel--Noack trapping theorem\label{sec:trapping_theorem}}
The Schlegel and Noack~\cite{Schlegel2015jfm} theorem, summarized in Theorem~\ref{th:trapping_theorem} below, provides necessary and sufficient conditions for the projected ROM in Eq.~\eqref{eq:Galerkin_model} to be globally stable by admitting a trapping region.
This theorem is necessary and sufficient for systems that exhibit effective nonlinearity, i.e., the system does not manifest invariant manifolds where there exists some $i$ such that $Q_{ijk}a_ja_k = 0$ for all time, for which a linear stability analysis must be adopted. In other words, systems that start in purely linear model subspaces, and remain in those subspaces, do not exhibit effective nonlinearity.
Fortunately, realistic fluid flows exhibit effective nonlinearity, although there are some subtleties we discuss in Section~\ref{sec:effective_nonlinearity}. 
In this case, we can always use the energy $K$ as a Lyapunov function for the trapping region. This is ideal, as finding a suitable Lyapunov function is often the most difficult task in stability analysis. 

A generic nonlinear system may exhibit multiple fixed points, limit cycles, and other equilibrium point behavior.  
However, any physical system should produce bounded trajectories, and the global stability property from the trapping theorem is agnostic to any \textit{local} stability properties.
This manuscript solely considers systems that are globally stable, or equivalently, long-term (ultimately) bounded, by virtue of exhibiting globally trapping regions. Long-term boundedness means that there exists some $T_0$ and $R_0$ such that $\|\bm{a}(t)\| < R_0$ for all $t > T_0$.
A trapping region encompasses an attractor or attracting set, which is typically defined as a set of the system phase space that many trajectories converge towards; this can be an equilibrium point, periodic trajectory, Lorenz's ``strange attractor'', or some other chaotic trajectory. Whenever it does not detract from the discussion, we omit the qualifiers ``globally'', ``monotonically'' and ``long-term'', as this is the only characterization of stability considered in the present work.
Examples of physical systems that are globally stable but do not exhibit a trapping region include Hamiltonian systems and systems that do not fit into the trapping theorem assumptions (examined further in Section~\ref{sec:effective_nonlinearity} and summarized in Fig.~\ref{fig:overview}). For instance, fully energy-preserving systems satisfy $\dot{K} = 0$, so trajectories represent shells of constant distance from the origin; these trajectories are globally bounded but no trapping region exists.
\vspace{0.05in}

\begin{mytheo}{Schlegel and Noack Trapping Theorem}{trapping_theorem}

This theorem provides necessary and sufficient conditions for energy-preserving, effectively nonlinear, quadratic systems to exhibit a trapping region $B(\bm{m},R_m)$, a ball centered at $\bm{m}$ with radius $R_m$.  
Outside this region the rate of change of energy $K$ is negative everywhere, producing a Lyapunov function that renders this system globally stable.  
Recentering the origin by an arbitrary constant vector $\bm{m}$, the energy may be expressed in terms of the shifted state vector $\bm{y}(t)=\bm{a}(t)-\bm{m}$ as
\begin{align}
\label{eq:K}
    K = \frac{1}{2}\bm{y}^T\bm{y}.
\end{align}
Taking a derivative and substituting in Eq.~\eqref{eq:Galerkin_model} produces
\begin{align}
\label{eq:Kdot}
\frac{d}{dt}K = \bm{y}^T\bm{A}^S\bm{y} + \bm{d}_m^T\bm{y},
\end{align}
\begin{align}\label{eq:def_AS_LS_dm}
    \bm{A}^S &= \bm{L}^S - \bm{m}^T\bm{Q}, \qquad \bm{L}^S = \frac{1}{2}(\bm{L} + \bm{L}^T), \quad \text{and}\quad
    \bm{d}_m = \bm{E} + \bm{L}\bm{m} + \bm{Q}\bm{m}\bm{m}.
\end{align}
$\bm{m}^T\bm{Q}$ refers to $m_iQ_{ijk}$ and $\bm{Q}\bm{m}\bm{m}$ to $Q_{ijk}m_jm_k$. 
The trapping theorem may now be stated as:

\textit{There exists a monotonically trapping region at least as small as the ball $B(\bm{m},R_m)$ if and only if the real, symmetric matrix $\bm{A}^S$ is negative definite$^*$\let\thefootnote\relax\footnotetext{$^*$ If a system is long-term bounded (not necessarily exhibiting a monotonically trapping region) and effectively nonlinear, only the existence of an $\bm{m}$ producing a negative \emph{semi}definite $\bm{A}^S$ is guaranteed.} with eigenvalues $\lambda_r \leq \cdots \leq \lambda_1 < 0$; the radius is then given by $R_m = \|\bm{d}_{m}\|/|\lambda_1|$. } 
\vspace{.15in}

In practice, the goal is then to find an origin $\bm{m}$ so that the matrix $\bm{A}^S$ is negative definite, guaranteeing a trapping region and global stability. 
Without effective nonlinearity, described at the beginning of Section~\ref{sec:trapping_theorem}, only the backwards direction holds; if we can find an $\bm{m}$ so that $\bm{A}^S$ is negative definite, the system exhibits a trapping region. However, such systems can be globally stable without admitting such an $\bm{m}$. Subsequently, the goal of Section~\ref{sec:methodology} is to use this theorem to define a constrained machine learning optimization that identifies a reduced-order model with a guaranteed trapping region. Even when the requirements of the trapping theorem are not fully satisfied, the algorithm results in Section~\ref{sec:results} indicate that this approach tends to produce models with improved stability properties.

\end{mytheo}

To understand the $R_m$ bound in Thm.~\ref{th:trapping_theorem}, we transform into eigenvector coordinates $\bm{z} = \bm{T}\bm{y}$, $\bm{h} = \bm{d}_m\bm{T}^T$, where $\bm{T}$ are the eigenvectors of $\bm{A}^S$.
Now Eq.~\eqref{eq:Kdot} becomes
\begin{align}
    \label{eq:ellipsoid_details}
    \frac{d}{dt}K = \sum_{i=1}^r h_iz_i + \lambda_i z_i^2 = \sum_{i=1}^r \lambda_i \left(z_i + \frac{h_i}{2\lambda_i}\right)^2 - \frac{h_i^2}{4\lambda_i}. 
\end{align}
We can see that the trapping region will be determined by the equation of the ellipsoid where $\dot{K} = 0$,
\begin{align}
    \label{eq:ellipsoid}
    1 = \sum_{i=1}^r \frac{1}{\alpha_i^2} \left(z_i + \frac{h_i}{2\lambda_i}\right)^2, \qquad \alpha_i = \frac{1}{2}\sqrt{\frac{1}{\lambda_i}\sum_{j=1}^r\frac{ h_j^2}{\lambda_j}} \leq \frac{1}{2|\lambda_1|}\|\bm{d}_m\|.
\end{align}
The origin at $\bm{y} = 0$ ($\bm{a} = \bm{m}$) lies on the ellipsoid, and in the worst case scenario lies at the tip of the major axis. Thus, to guarantee that a ball centered at this origin entirely contains this region, we estimate $R_m$ as twice the size of the largest possible value of the half-axes $\alpha_i$.
Note that our definition of $\alpha_i$ differs from Schlegel and Noack~\cite{Schlegel2015jfm}; we believe that there is a minor typo in their Eq. 3.14. Fortunately, the only consequence is a change in the estimate of $R_m$. 
Lastly, recall that long-term bounded (not necessarily exhibiting a monotonically trapping region) and effectively nonlinear systems only guarantee an $\bm{m}$ exists such that $\bm{A}^S$ is negative semidefinite. In the case of mixed zero and nonzero eigenvalues, the ellipsoid becomes a paraboloid. The infinite extent of the paraboloid precludes a monotonic trapping region but not other forms of global stability. This edge case is not further discussed because in practice (numerically) there is no chance of arriving at an eigenvalue of exactly zero.

\subsection{Model truncation, effective nonlinearity, and closure models \label{sec:effective_nonlinearity}}
Before implementing the trapping theorem into system identification, we investigate the circumstances under which truncated projection-based models will exhibit effective nonlinearity; the reader may skip this section if the subtleties of the trapping theorem are not of interest, although the discussion here is pertinent to Section~\ref{sec:results_burgers}.
Effectively nonlinear dynamics are ideal because they can be decisively classified as globally stable or not, requiring no additional stability analysis.  
To proceed, consider a Fourier-Galerkin model of Burgers' equation derived from the Fourier expansion $u(x,t) = \sum a_k(t)e^{ikx}$, and further examined in Section~\ref{sec:results_burgers},
\medmuskip=0mu
\thickmuskip=0mu
\thinmuskip=0mu
\begin{align}
\label{eq:burgers_cascade}
 \dot{u} = -u\partial_x u + \nu \partial_{xx}u \quad \Longrightarrow \quad
    \dot{a}_k = -\nu k^2a_k - \sum_{\ell=-\infty}^{\infty} i \ell a_{\ell} a_{k - \ell} \quad \Longrightarrow \quad
    \dot{K} = -\nu\sum_{k=-\infty}^\infty k^2 a_k^2 - \sum_{k,\ell=-\infty}^{\infty} i \ell a_{\ell} a_{k - \ell}a_k.
\end{align}
\medmuskip=4mu
\thickmuskip=4mu
\thinmuskip=4mu
The particular ``triadic'' structure of the nonlinear term in the spectral domain, where the only nonzero terms acting on $a_k$ are those whose wavenumbers sum to $k$, is identical to that arising in isotropic turbulence~\cite{Tennekes1972book}.
The triadic term in $\dot{K}$ transfers energy between length scales. 
Since the viscous term scales with $k^2$, energy is most effectively dissipated at the smallest scales; the combination of the two terms leads to the traditional energy cascade in which energy flows ``downhill'' from larger to smaller scales.
This description implies that heavily truncating the Galerkin system leads to under-resolving the dissipation rate and a closure scheme may be required to re-introduce the full dissipation. 
Towards this goal, modern sparse regression and deep learning methods have been used to produce new closures for fluid models~\cite{ling2016reynolds,san2018neural,pan2018data,Duraisamy2019arfm,Maulik2019jfm,beetham2020formulating}. 
While the traditional explanation for unstable Galerkin models derives from these truncated dissipative scales, increasingly there are alternate explanations including fundamental numerical issues with the Galerkin framework (potentially resolved in a Petrov-Galerkin framework)~\cite{grimberg2020stability} and the Kolmogorov width issues of linear subspaces more generally~\cite{lee2020model}. If true, this is probably good news for (incompressible, dissipationless) Hall-MHD, where the conservation of energy and the cross, magnetic, and generalized helicities leads to direct, inverse, and even bidirectional cascades~\cite{pouquet2019helicity}. Interestingly, the notion of effective nonlinearity appears to be another approach from which we can attempt to resolve these disagreements about the sources of ROM instability.

To proceed with this theme, we show that the triadic structure of the model has repercussions for the presence of effective nonlinearity. 
Consider the truncated model
\begin{align}
    \dot{a}_k = -\nu k^2a_k - \sum_{\ell=-r}^{r} i\ell a_{\ell} a_{k - \ell}, \qquad k \in \{1,...,r\}
\end{align}
with the initial condition $a_{j} = 1$ for any $j \in \{\pm(\frac{r}{2}+1), \pm(\frac{r}{2}+2),..., \pm r\}$, and $a_k = 0$, $k \neq j$. For simplicity we have assumed $r$ is divisible by two. In this case the system has $r$ invariant 1D subspaces for which 
\begin{align}
\label{eq:subspaces}
    \dot{a}_{j} = -\nu j^2a_{j}.
\end{align}
These invariant linear subspaces exist because higher wavenumber modes that \emph{could} interact to transfer energy between coefficients have been discarded. 
In other words, \textit{Fourier-Galerkin models with finite truncation do not exhibit effective nonlinearity}. 
In contrast, POD-Galerkin models weakly break the triadic structure of the nonlinearity~\cite{couplet2003intermodal}, and therefore in general will weakly satisfy the trapping theorem criteria for effective nonlinearity, to the extent that they differ from the Fourier modes because of inhomogeneity in the system. 
There are also modern ROMs which attempt to retain the full dissipation by utilizing bases that intentionally mix length scales~\cite{balajewicz2012stabilization} -- these types of models should be more likely to satisfy effective nonlinearity. 
Lastly, numerical errors appear to weakly restore effective nonlinearity, since errors break any triadic structure. 
Proceeding with this analysis is complicated because the numerical errors also weakly break our foundational assumption that $Q_{ijk}$ is energy-preserving. Future investigations should be pursued to explore relationships between effective nonlinearity, the energy cascade, and closure models that reintroduce stabilizing dissipation to truncated models.

It is difficult to quantify ``how close'' a model is to exhibiting effective nonlinearity, since a lack of effective nonlinearity $Q_{ijk} a_j a_k = 0$ must hold for all time, for any $i$, and for any valid system trajectory. 
However, for an orthonormal set of temporal modes, and assuming there exists at least one index $i$ such that $Q_{ijj} \neq 0$, we propose quantifying the average strength of model effective nonlinearity through the metric
\begin{align}
\label{eq:effective_nonlinearity_strength}
    S_e = \frac{\min_i |Q_{ijk}\overline{a_ja_k}|}{\max_i |Q_{ijk}\overline{a_ja_k}|} = \frac{\min_i |Q_{ijj}|}{\max_i |Q_{ijj}|}.
\end{align}
The bar in $\overline{a_ja_k}$ denotes a temporal average. 
We will show in Section~\ref{sec:results_burgers} that in system identification a lack of effective nonlinearity is not a terrible loss. Our trapping SINDy algorithm in Section~\ref{sec:methodology} minimizes $\dot{K}$ whether or not a negative definite $\bm{A}^S$ can be realized. 
However, without additional stability analysis, such models are no longer provably stable for any initial condition. 
Although Eq.~\eqref{eq:subspaces} is a linearly stable system, this is not guaranteed for more general fluid models than the simple Burgers' equation considered here. 

\subsection{Constraints in model reduction and system identification\label{sec:constraints}}
Before moving on to system identification, it is worth noting that enforcing these types of existence-based stability conditions is subtle. There are modern techniques to implement implicit constraints of the form
\begin{align}
\label{eq:general_implicit_constraint}
    \mathcal{C}_i(\dot{\bm{a}},\bm{a},t,...) = 0,\,\,\,\,\,\,\,\,\, i \in \{1,2,...\}
\end{align}
into both model reduction~\cite{lee2019deep,schein2020preserving} and system identification~\cite{loiseau2018constrained,kolter2019learning,champion2020unified,srivastava2021generalizable}. Precisely in this way, the energy-preserving constraint in Eq.~\eqref{eq:energy_preserving_nonlinearity} is cast as an affine version of Eq.~\eqref{eq:general_implicit_constraint} in our optimization in Section~\ref{sec:methodology}. 

However, enforcing stability in quadratic energy-preserving models is more complicated than Eq.~\eqref{eq:general_implicit_constraint}. To see this, note that there a few different circumstances under which we might want to promote stability. If the true $\bm{A}^S$ \textit{and} the optimal $\bm{m}$ are known, we can simply constrain the coefficients in Eq.~\eqref{eq:Galerkin_model} to produce this known negative definite $\bm{A}^S$. This would imply that we already know the optimally-shifted eigenvalues of the system and an $\bm{m}$ that produces these negative eigenvalues; if this is the case, so much information about the system of ODEs is already known that machine learning methods are likely unnecessary. 

But far more interesting are the cases in which 1) the underlying system is known to be globally stable and effectively nonlinear, so we want to find the ``correct'' $\bm{m}$ and corresponding $\bm{A}^S$, or 2) it is not known if any $\bm{m}$ exists such that $\bm{A}^S$ is negative definite. In system identification, either of these cases can be addressed by searching for a model that both optimally fits the data and is globally stable. In this context, we adopt a mixed approach in the next section where we enforce the energy-preserving constraint and then separately bias the optimization towards models with a trapping region. 
This technique is a significant methodological extension because we can no longer rely on constraints of the form in Eq.~\eqref{eq:general_implicit_constraint}.

\section{Trapping SINDy algorithm \label{sec:methodology}}
We now describe how to incorporate the trapping theorem of Schlegel and Noack~\cite{Schlegel2015jfm} into data-driven model identification, specifically for the SINDy algorithm. 
Before describing the modified algorithm in Sec.~\ref{sec:sindy_trapping}, we first present the standard SINDy algorithm~\cite{Brunton2016pnas} along with a recent variant that incorporates explicit constraints~\cite{loiseau2018constrained,champion2020unified}.  
We then build on this framework to incorporate the Schlegel--Noack trapping theorem. 

\subsection{Standard and constrained SINDy algorithms\label{sec:sindy_vanilla}}
The goal of system identification is to identify a system of ODEs or PDEs that describe how a given data set evolves dynamically in time.  
The SINDy algorithm~\cite{Brunton2016pnas} identifies sparse,  parsimonious models that remain interpretable and avoid the overfitting issues that are common in this field. 
As in Loiseau et al.~\cite{loiseau2018constrained}, we develop SINDy models for the dynamics of $\bm{a}$, representing the coefficients or amplitudes of a modal Galerkin expansion in Eq.~\eqref{eq:galerkin_expansion}.  
We assume that the dynamics of $\bm{a}$ will be described as a sparse linear combination of elements from a library $\bm{\Theta}$ containing candidate terms such as:
\begin{align}
\label{eq:SINDyExpansion}
    \frac{d}{dt}{\bm{a}} \approx \bm{\Theta}(\bm{a})\bm{\Xi},\qquad
\bm{\Theta}(\bm{a}) = 
\begin{bmatrix} 
~~\vline&\vline & \vline\\
~~\bm{1}&\hspace{-.025in}\bm{a} & \hspace{-.025in}\bm{a}\otimes\bm{a} & \hspace{-.025in}\\
~~\vline &\vline & \vline 
\end{bmatrix}. 
\end{align}
Here $\bm{a}\otimes\bm{a}$ contains all combinations of $a_ia_j$ without duplicates. 
The $\bm{\Theta}$ matrix may contain any desired candidate terms, but in this work we consider only terms up to quadratic polynomials in $\bm{a}$ because we are searching for energy-preserving quadratic models.
The expressions in Eq.~\eqref{eq:SINDyExpansion} are typically evaluated on a data matrix $\bm{X}$ obtained from time-series data of the state, $\bm{a}(t_1),\bm{a}(t_2),...,\bm{a}(t_M)$:
\begin{eqnarray}
\bm{X} = \overset{\text{\normalsize state}}{\left.\overrightarrow{\overset{~~}{\begin{bmatrix}
a_1(t_1) & a_2(t_1) & \cdots & a_r(t_1)\\
a_1(t_2) & a_2(t_2) & \cdots & a_r(t_2)\\
\vdots & \vdots & \ddots & \vdots \\
a_1(t_M) & a_2(t_M) & \cdots & a_r(t_M)
\end{bmatrix}}}\right\downarrow}\begin{rotate}{270}\hspace{-.125in}time~~\end{rotate} \hspace{.125in}
\label{Eq:DataMatrix}.
\end{eqnarray}
A matrix of derivatives in time, $\dot{\bm{X}}$, is defined similarly and can be numerically computed from $\bm{X}$. 
In this case, Eq.~\eqref{eq:SINDyExpansion} becomes $\dot{\bm{X}} = \bm{\Theta}(\bm{X})\bm{\Xi}$. 
The goal of SINDy is to determine a sparse matrix of coefficients $\bm{\Xi} = \begin{bmatrix}\bm{\xi}_1 &\bm{\xi}_2 & \cdots & \bm{\xi}_r\end{bmatrix}$, also written in vectorized form as 
\begin{align}
\label{eq:vectorized_xi}
   \bm{\Xi}[:] = \bm{\xi} =  [\xi^{a_1}_1, \ldots, \xi^{a_r}_1,\xi^{a_2}_2 ,\ldots, \xi^{a_r}_2,\ldots, \xi^{a_1}_N, \ldots, \xi^{a_r}_N], 
\end{align} where $N$ is the number of candidate functions and $r$ is the state space size; nonzero elements in each column $\bm{\xi}_j$ indicate which terms are active in the dynamics of $\dot{a}_j(t)$. 
The matrix of coefficients $\bm{\Xi}$ is determined via the following sparse optimization problem:
\begin{align}
\label{eq:optimization_vanilla}
\argmin_{\bm{\xi}}&\left[ \frac{1}{2}\|\bm{\Theta}\bm{\xi}-\dot{\bm{X}}\|^2 + \lambda \|\bm{\xi}\|_0\right].
\end{align}
We deviate from the typical SINDy definitions by explicitly formulating the problem in terms of the vectorized $\bm{\xi} \in \mathbb{R}^{rN}$, $\bm{\Theta}(\bm{X}) \in \mathbb{R}^{rM\times rN}$, and $\dot{\bm{X}} \in \mathbb{R}^{rM}$. The first term in the SINDy optimization problem in Eq.~\eqref{eq:optimization_vanilla} fits a system of ODEs $\bm{\Theta}\bm{\xi}$ to the given data in $\dot{\bm{X}}$.
The $\|\bm{\xi}\|_0$ term counts the number of nonzero elements of $\bm{\Xi}$; however, it is not technically a norm and leads to a non-convex optimization, so several convex relaxations have been proposed~\cite{Brunton2016pnas,Rudy2017sciadv,champion2020unified}.

Since the original SINDy algorithm, Loiseau et al.~\cite{loiseau2018constrained} introduced an extension to directly enforce constraints on the coefficients in $\bm{\Xi}$. 
In particular, they enforced energy-preserving, skew-symmetry constraints on the quadratic terms for incompressible fluid flows, demonstrating improved model performance over standard Galerkin projection. 
The quadratic library in Eq.~\eqref{eq:SINDyExpansion} has ${N = \frac{1}{2}(r^2+3r)+1}$ terms. 
With the energy-preserving structure, it can be shown that the number of constraints is ${p = r(r+1)(r+2)/6}$ and therefore the number of free parameters is $rN - p = 2p$. This constraint is encoded as $\bm{C}\bm{\xi} = \bm{d}$, $\bm{C} \in \mathbb{R}^{p\times rN}$, $\bm{d} \in \mathbb{R}^{p}$, and the constrained SINDy algorithm solves the following minimization,
\begin{align}
\label{eq:optimization_constrained}
\argmin_{\bm{\xi}}&\left[ \frac{1}{2}\|\bm{\Theta}\bm{\xi}-\dot{\bm{X}}\|^2_2 + \lambda \|\bm{\xi}\|_1+ \delta_0(\bm{C}\bm{\xi} - \bm{d})\right].
\end{align}
In general we can use nonconvex regularizers that promote sparsity in $\bm{\xi}$, but the trapping SINDy modifications below require a convex regularizer such as the $L^1$ norm. The third term $\delta_0$ is an indicator function that encodes 
the constraint $\bm{C}\bm{\xi} = \bm{d}$, guaranteeing the energy-preserving structure in the quadratic nonlinearity is retained in the identified model. There are also variants of the constrained SINDy objective function in Eq.~\eqref{eq:optimization_constrained} that utilize sparse relaxed regularized regression (SR3) in order to improve performance~\cite{zheng2019unified,champion2020unified}. 

\subsection{Proposed trapping SINDy algorithm}\label{sec:sindy_trapping}
Model constraints in system identification, such as global conservation laws or other physical considerations, often result in improved models, but do not generally guarantee global stability. 
Here, we will additionally promote globally stable models that exhibit a monotonically trapping region. 
Recall from Thm.~\ref{th:trapping_theorem} that $\bm{m}$ is an arbitrary, constant vector, of the same state size as $\bm{a}$, that specifies the center of a possible trapping region. 
Stability promotion is then achieved by jointly determining the sparse model coefficients $\bm{\Xi}$ and state vector $\bm{m}$ such that $\bm{A}^S$ from Eq.~\eqref{eq:def_AS_LS_dm} is negative definite. 

To proceed with our trapping SINDy formulation, we must relate the model coefficients in $\bm{\xi}$ to the matrix $\bm{A}^S$ appearing in the trapping theorem. 
We first define the projection operators ${\bm{P}^L\in \mathbb{R}^{r \times r \times rN}}$, ${\bm{P}^Q\in \mathbb{R}^{r \times r \times r \times rN}}$, and ${\bm{P}\in \mathbb{R}^{r \times r \times rN}}$. 
The operator $\bm{P}^L$ projects out the symmetric part of the linear coefficients through $\bm{L}^S = \bm{P}^L\bm{\xi}$. The same is true for the quadratic coefficients, $\bm{Q} = \bm{P}^Q\bm{\xi}$. The operator ${\bm{P} = \bm{P}^L - \bm{m}^T\bm{P}^Q}$ provides a concise representation of $\bm{A}^S$ through the following equation:
\begin{align} \label{eq:W_def}
\bm{A}^S = \bm{L}^S - \bm{m}^T\bm{Q} = \bm{P}\bm{\xi} = (\bm{P}^L-\bm{m}^T\bm{P}^Q)\bm{\xi}.
\end{align}

We now phrase a tentative version of the trapping SINDy optimization problem, in analogy to the constrained SINDy optimization in Eq.~\eqref{eq:optimization_constrained}, that incorporates an additional loss term to reduce the maximal (most positive) eigenvalue $\lambda_1$ of the real, symmetric matrix $\bm{A}^S$:
\begin{align}
\label{eq:optimization_alternate}
\argmin_{\bm{\xi},\bm{m}}&\left[ \frac{1}{2}\|\bm{\Theta}\bm{\xi}-\dot{\bm{X}}\|^2_2 + \lambda \|\bm{\xi}\|_1+ \delta_0(\bm{C}\bm{\xi} - \bm{d}) + \frac{1}{\eta}\lambda_1\right].
\end{align}
Although $\lambda_1$ is a convex function of the matrix elements~\cite{overton1988minimizing}, $(\bm{P}^L - \bm{m}^T\bm{P}^Q)\bm{\xi}$ is not affine in $\bm{\xi}' = [\bm{\xi}, \bm{m}]$. The result is that this new term is not convex, but {\it convex composite}. It is possible to approximately solve this problem with a variable projection technique, where we essentially treat $\bm{\xi}$ and $\bm{m}$ as independent, solve the convex problem in $\bm{\xi}$, and then substitute $\bm{\xi}^*$, the solution at each iteration, into the optimization for $\bm{m}$. 
In practice this algorithm performs fairly well, although the convergence properties are unclear.  
Eq.~\eqref{eq:optimization_alternate} is also amenable to other approaches, such as Gauss-Newton~\cite{burke1995gauss} or the prox-linear algorithm~\cite{drusvyatskiy2019efficiency}, because $\lambda_1$ is a convex function and $\bm{P}\bm{\xi}$ is smooth in $\bm{m}$ and $\bm{\xi}$. Although we institute a modified algorithm below, these convex-composite approaches are a promising future direction for effectively solving this nonconvex optimization problem. 

In order to produce an algorithm with better performance and better understood convergence properties, we adopt a relax-and-split approach~\cite{zheng2020relax}, similar to the approach taken in Champion et al.~\cite{champion2020unified}.
We introduce an auxiliary variable $\bm{A}$ that represents the projection of $\bm{A}^S=\bm{P}\bm{\xi}$ onto the space of negative definite matrices, and introduce two new terms in the optimization:
\begin{align}
\label{eq:optimization_wAm}
\argmin_{\bm{\xi},\bm{m},\bm{A}}&\left[ \frac{1}{2}\|\bm{\Theta}\bm{\xi}-\dot{\bm{X}}\|^2_2 + \lambda \|\bm{\xi}\|_1+ \delta_0(\bm{C}\bm{\xi} - \bm{d}) +
\frac{1}{2\eta}\|\bm{P}\bm{\xi}-\bm{A}\|^2_2 +\delta_{\mathcal{I}}(\bm{\Lambda})\right].
\end{align}
The new least-squares term enforces a ``soft'' constraint (or bias) towards $\bm{A}^S=\bm{P}\bm{\xi}$ being negative definite by minimizing the difference between $\bm{P}\bm{\xi}$ and its projection into the space of negative definite matrices. 
The auxiliary variable $\bm{A}$ is updated to approximate $\bm{A}^S=\bm{P}\bm{\xi}$, and then, through the $\delta_{\mathcal{I}}$ term, enforced to be negative definite by requiring that the diagonalized matrix $\bm{\Lambda} = \bm{V}^{-1}\bm{A}\bm{V}$ lies in
$\mathcal{I} = (-\infty, -\gamma\,]$, $\gamma > 0$. 
Directly enforcing $\bm{P}\bm{\xi}$ to be negative definite tends to wreck the model fit to the data. 
Instead, the auxiliary variable $\bm{A}$ in Eq.~\eqref{eq:optimization_wAm} allows the algorithm to accurately fit the data with $\bm{\xi}$ and then relax the coefficients towards a negative definite $\bm{A}^S$ to promote global stability. 

This flexible formulation also allows $\bm{A}$, our proxy for the projection of $\bm{P}\bm{\xi}$ onto the space of negative definite matrices, to vary, and therefore fit the particular eigenvalues of the system in question. 
In other words, the proposed approach pushes $\bm{A}^S$ into the space of negative definite matrices in $\mathbb{R}^{r\times r}$ with minimal assumptions about the eigenvalues, only assuming that they are negative. Contrast our algorithm to a more restrictive approach that prescribes an $\bm{A}$, meaning we already know a set of negative eigenvalues of $\bm{P\xi}$ that is compatible with the data. 
A description of each of the hyperparameters $\lambda$, $\eta$, and $\gamma$, is provided in Table~\ref{tab:hyperparams}.
Note that Eq.~\eqref{eq:optimization_wAm} is not convex in $\bm{A}$, and this is the most challenging aspect of this formalism. 

Now that we have defined our problem in Eq.~\eqref{eq:optimization_wAm}, we need to solve it. If we denote the convex part of the optimization,
\begin{align}
    \label{eq:convex_optimization_piece}
F(\bm{\xi},\bm{m},\bm{A}) = \|\bm{\Theta}\bm{\xi}-\dot{\bm{X}}\|^2_2/2 + \lambda \|\bm{\xi}\|_1+ \delta_0(\bm{C}\bm{\xi} - \bm{d}) +
\|\bm{P}\bm{\xi}-\bm{A}\|^2_2/2\eta,
\end{align}
and fix initial guesses for $\bm{m}$ and $\bm{A}$, then we can define the solution vector $\bm{\xi}^*$ through
\begin{align}
    \label{eq:optimization_xi}
    \bm{\xi}^* = \argmin_{\bm{\xi}}\left[F(\bm{\xi}, \bm{m}, \bm{A})\right].
\end{align}
\begin{table}[t]
\centering
\begin{tabular}{ |>{\columncolor[gray]{0.85}}p{0.15cm}|p{15cm}|  }
 \hline
 \rowcolor{gray!30} \multicolumn{2}{|c|}{Trapping SINDy hyperparameters} \\
 \hline
 $\lambda$ &
 Specifies the strength of sparsity-promotion through the regularizer $R(\bm{\xi})$. $\lambda=0$ already works well for low-dimensional systems because the $\|\bm{P}\bm{\xi} - \bm{A}\|^2_2$ term promotes stability. \\
 \hline
 $\eta$ & Specifies how strongly to push the algorithm towards models with negative definite $\bm{A}^S$. If $\eta \gg 1$, $\bm{\xi}^*$ is unaffected by the minimization over $\bm{m}$. If $\eta \ll 1$, the problem is increasingly nonconvex. \\
 \hline
 $\gamma$ 
& Determines how far to push the eigenvalues of $\bm{A}^S$ towards being negative definite. Typically $\gamma \lesssim 0.1$ works for a variety of problems regardless of the true eigenvalues of $\bm{A}^S$. \\
 \hline
\end{tabular}
\caption{Description of the trapping SINDy hyperparameters. }
\label{tab:hyperparams}
\vspace{-0.2in}
\end{table}
If $\lambda = 0$, $\bm{\xi}^*$ is structurally identical to the $\bm{\xi}^*$ in Champion et al.~\cite{champion2020unified}:
\begin{align}
\label{eq:H}
    \bm{H} &= (\bm{\Theta}^T\bm{\Theta} + \frac{1}{\eta}\bm{P}^T\bm{P})^{-1}, \\
\label{eq:w_update}
    \bm{\xi}^* &= \bm{H}\left[\bm{I} - \bm{C}^T(\bm{CHC}^T)^{-1}\bm{CH}\right]\left[\bm{\Theta}^T\dot{\bm{X}} + \frac{1}{\eta}\bm{P}^T\bm{A}\right] + \bm{H}\bm{C}^T(\bm{CHC}^T)^{-1}\bm{d}.
\end{align}
$\bm{H}$ is positive definite, $\bm{I}$ is the identity matrix, and $\bm{C}\bm{\xi}^* = \bm{d}$ can be verified using Eq.~\eqref{eq:w_update}. The minimization over $\bm{\xi}$ with $\lambda \neq 0$ is still convex but not analytically tractable as in Eq.~\eqref{eq:w_update}. Since it is convex, it can be solved with standard convex optimization libraries such as CVXPY~\cite{diamond2016cvxpy}. It can also be shown to reduce to a constrained quadratic program over the unit box with a positive semidefinite cost matrix. A barrier to this route is that typical numerical solvers either assume that the quadratic cost matrix is sparse or positive definite. Neither assumption is true here. 

Now that we have solved the minimization over $\bm{\xi}$, we can use prox-gradient descent on $(\bm{m}, \bm{A})$; each algorithm iteration we alternate between solving for $\bm{\xi}^*$ and solving for $(\bm{m}^*, \bm{A}^*)$.
Again, we can think about this problem as a variable projection~\cite{aravkin2016variable,zhang2020offline}, which is a value function optimization over the remaining variables $(\bm{m},\bm{A})$. 
To make this viewpoint more precise, we define  
\begin{align}
\widetilde F(\bm{m},\bm{A}) = F(\bm{\xi}^*,\bm{m},\bm{A}), 
\end{align}
The problem we want to solve is now written more simply as 
\[
\argmin_{\bm{m},\bm{A}} \left[\widetilde F(\bm{m},\bm{A}) + \delta_{ \mathcal{I}}(\bm{\Lambda})\right]. 
\]
We apply prox-gradient descent to this nonconvex problem, so that
\begin{align}
\label{eq:mA_update}
\bm{m}^* = \bm{m} - \alpha_m\nabla_{\bm m} \widetilde F(\bm{m},\bm{A}),  \qquad
\bm{A}^* = \text{proj}_{\mathcal{I}}\left[\bm{A} - \alpha_A\nabla_{\bm A} \widetilde F(\bm{m},\bm{A})\right],
\end{align}
where $\alpha_m$ and $\alpha_A$ are step sizes. 
All that remains is to compute the gradients of the value function $\widetilde F$,
\begin{align}
\nabla_{\bm A} \widetilde F(\bm{m},\bm{A}) = \frac{1}{\eta}(\bm{A}-\bm{P}\bm{\xi}^*), \qquad
\nabla_{\bm m} \widetilde F(\bm{m},\bm{A}) = \frac{1}{\eta}\bm{P}^Q\bm{\xi}^*(\bm{A}-\bm{P}\bm{\xi}^*).
\end{align}
These are Lipschitz continuous functions with Lipschitz constants $L_A$, $L_m$ satisfying
\begin{align}
    \label{eq:lipshitz_m}
    \alpha_A \leq \frac{1}{L_A} \leq \eta, \qquad
    \alpha_m \leq \frac{1}{L_m} \leq \frac{\eta}{\|(\bm{P}^Q\bm{\xi}^*)_{ijk}(\bm{P}^Q\bm{\xi}^*)_{ljk}\|_\text{F}},
\end{align}
in order for guaranteed convergence of fixed step-size, prox-gradient descent~\cite{attouch2013convergence}. While the denominator in Eq.~\eqref{eq:lipshitz_m} varies with the update in $\bm{\xi}$, in practice, one can reduce $\alpha_m$ until convergence is found. 

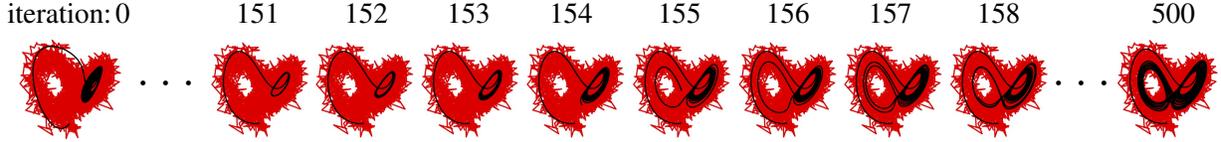
\begin{figure}[t]
\vspace{0.2in}
\begin{flushright}    \begin{overpic}[width=.97\linewidth]{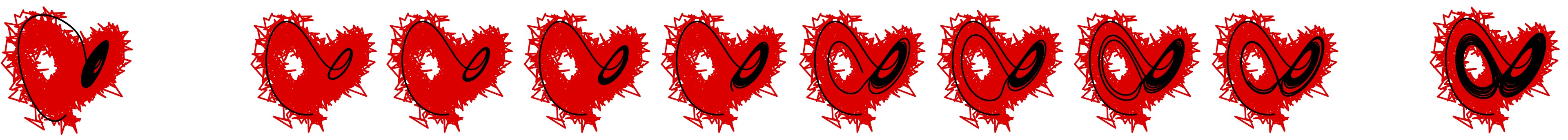}
    \put(-1, 10){iteration:}
    \LARGE{
    \put(9.75, 4){$\bm{\cdots}$}
    \put(85.75, 4){$\bm{\cdots}$}}
    \normalsize
    \put(8, 10){0}
    \put(18, 10){151}
    \put(27, 10){152}
    \put(35.5, 10){153}
    \put(44, 10){154}
    \put(53, 10){155}
    \put(62, 10){156}
    \put(70.5, 10){157}
    \put(79.5, 10){158}
    \put(94, 10){500}
    \end{overpic}
    \end{flushright}
    \vspace{-0.225in}
    \caption{Illustration of trapping SINDy progress on noisy Lorenz data. The minimization results in the transition from a poor initial guess to identification of the correct attractor dynamics.}
    \label{fig:SINDy_progress}
    \vspace{-0.2in}
\end{figure}

The full trapping SINDy optimization is illustrated in Algorithm~\ref{algo}.
\begin{algorithm*}[t]
	\caption{Trapping  SINDy} \label{algo}
	\begin{algorithmic}[1]
		\INPUT{Numerical data $\dot{\bm{X}}$ and optional initial guesses for $\bm{m}$ and $\bm{A}$.}
		\OUTPUT{Optimal model coefficients $\bm{\xi}^*$ and shift vector $\bm{m}^*$}.  
		\Procedure{SINDy}{$\dot{\bm{X}}$, $\lambda$, $\eta$, $\gamma$}
		\State Construct matrices $\bm{\Theta}(\bm{X})$, $\bm{P}$, $\bm{C}$, and $\bm{d}$.
		\Comment{Initialize variables}
		\State $\textbf{while}\,\,\,\, |\bm{\xi}_k - \bm{\xi}_{k+1}| > \epsilon_\text{tol}^\xi  \text{ and } |\bm{m}_k - \bm{m}_{k+1}| > \epsilon^m_\text{tol}$  \Comment{Begin iteration loop}
		\State \quad $\bm{\xi}_{k+1} \Longleftarrow \argmin_{\bm{\xi}_k}\left[F(\bm{\xi}_k,\bm{m}_k,\bm{A}_k)\right]$, \Comment{Convex minimization for $\bm{\xi}_{k+1}$}
		\State \quad $\bm{V}_{k+1}\bm{\Lambda}_{k+1}(\bm{V}_{k+1})^{-1} \Longleftarrow \bm{A}_k - \alpha_A\nabla_{\bm A} \widetilde F(\bm{m},\bm{A})|_{m_k,A_k}$,
		\Comment{Prox-gradient step for $\bm{A}$}
		\State \quad 
		$\bm{A}_{k+1} \Longleftarrow \bm{V}_{k+1}\text{proj}_{\mathcal{I}}\left[\bm{\Lambda}_{k+1} \right](\bm{V}_{k+1})^{-1}$,
		\Comment{Project $\bm{A}$ into $\mathcal{I}$, rotate into $\bm{P\xi}$ basis}
		\State \quad $\bm{m}_{k+1} \Longleftarrow \bm{m}_k - \alpha_m\nabla_{\bm m} \widetilde F(\bm{m},\bm{A})|_{m_k,A_k}$,
		\Comment{Prox-gradient step for $\bm{m}$}
		\EndProcedure
	\end{algorithmic}
Note that inequalities~\eqref{eq:lipshitz_m} should be satisfied, and there tends to be a sweet spot for $\eta$. It is often useful to start with $\eta \gg 1$ and then reduce $\eta$ until the model coefficients are significantly affected. 
\end{algorithm*}
$\epsilon_\text{tol}^\xi$ and $\epsilon_\text{tol}^m$ are convergence tolerances. The $\bm{V}_{k+1}$ are the eigenvectors of $\bm{P}\bm{\xi}_{k+1}$ and are used to transform $\bm{A}$ into the same basis as $\bm{P}\bm{\xi}_{k+1}$. 
An example of the algorithm iterating on noisy data from the chaotic Lorenz system is shown in Fig.~\ref{fig:SINDy_progress}, demonstrating how the algorithm transitions from a poor initial guess that decays to a fixed point to a stable model converging to the correct attractor. 
We also implement an optional FISTA method~\cite{beck2009fast,nesterov2013gradient} for reducing the convergence time in the $(\bm{m},\bm{A})$ optimization. Algorithm~\ref{algo} is computationally intensive, but it can be parallelized for speed in future work, following other SINDy variants~\cite{kaheman2020sindy}. Initial guesses for $\bm{m}$ and $\bm{A}$ also facilitate continuation of previous optimization runs. Along with these methods, we also implement the $\lambda_1$ variant of the trapping algorithm in Eq.~\eqref{eq:optimization_alternate} in the open-source PySINDy code~\cite{silva2020pysindy}.

A key insight to the trapping algorithm is that the energy-preserving constraint $\bm{C}\bm{\xi} = \bm{d}$ is non-negotiable. Although in practice small errors in $\bm{C}\bm{\xi} = \bm{d}$ do not significantly affect the optimization problem, the $\|\bm{P}\bm{\xi} - \bm{A}\|^2_2$ term in the optimization loses its physical interpretation if the coefficients are not exactly energy-preserving. %
Thus, the goal is to satisfy $\bm{C}\bm{\xi} = \bm{d}$ \textit{exactly}, and then to push a potential model towards a more refined model that exhibits a trapping region, potentially at the expense of the fit to the data (this can also mitigate overfitting). 
There tends to be a ``sweet spot'' regime for the value of $\eta$. If $\eta^{-1}\|\bm{P}\bm{\xi}-\bm{A}\|^2_2 \ll \|\bm{\Theta}\bm{\xi} - \dot{\bm{X}}\|^2_2$, then $\bm{\xi}^*$ is essentially unaffected by the minimizations over $\bm{m}$ and $\bm{A}$. In practice, this means that poor initial guesses for $\bm{\xi}^*$ do not improve as the full optimization problem is solved. In the opposite extreme, $\eta^{-1}\|\bm{P}\bm{\xi}-\bm{A}\|^2_2 \gg \|\bm{\Theta}\bm{\xi} - \dot{\bm{X}}\|^2_2$, the optimization in Eq.~\eqref{eq:optimization_wAm} is increasingly nonconvex and potentially pulls $\bm{\xi}^*$ far away from the data. Finding the $\eta$ regime where updating $\bm{m}$ perturbs $\bm{\xi}^*$, instead of leaving $\bm{\xi}^*$ unaffected or mangled, requires scanning $\eta$. 

\section{\label{sec:results}Results}
We now investigate the utility of our trapping SINDy algorithm to identify stable, sparse, nonlinear models for a number of canonical systems. 
These examples illustrate that it is possible to both effectively discover stable models that exhibit trapping regions and improve the discovery of systems that do not satisfy Thm.~\ref{th:trapping_theorem} or the requirement of effective nonlinearity. 
For each system, we train SINDy on a single trajectory with a random initial condition and evaluate the model on a different trajectory of the same temporal duration with a new random initial condition. 
It is difficult to quantity model performance for chaotic systems, such as the Lorenz system, where lobe-switching is extremely sensitive to initial conditions and the coefficients of the identified model, and for systems with transients, for which the precise timing of instability must be matched to achieve the correct phase. 
A reasonable definition for the model quality, for models with closed forms, is the relative Frobenius error in the model coefficients,
\begin{align}
\label{eq:model_error}
    \text{E}_\text{m} = \frac{\|\bm{\Xi}_\text{True}- \bm{\Xi}_\text{SINDy}\|_\text{F}}{\|\bm{\Xi}_\text{True}\|_\text{F}}.
\end{align}
When appropriate, we also report a far more demanding relative prediction error,
\begin{align}
\label{eq:prediction_error}
    \text{E}_\text{pred} = \frac{\|\bm{X}_\text{True} - \bm{X}_\text{SINDy}\|_F}{ \|\bm{X}_\text{True}\|_F}.
\end{align}

Table~\ref{tab:results_summary} summarizes the sampling, hyperparameters, and identified trapping regions for each example discussed in Sections~\ref{sec:results_meanfield}--\ref{sec:results_vonKarman}. Table~\ref{tab:results_summary} is intended to be instructive rather than exhaustive. For clarity, the training and testing trajectories used to generate this table do not have added noise, although Fourier modes from the Burgers' Equation and POD modes from the Von K\`arm\`an street are obtained from direct numerical simulation (DNS), and subsequently contain minor numerical noise; the performance on noisy data will be explored further in Sec.~\ref{sec:results_Lorenz}. 
To compare trapping region sizes $R_m$ across different examples, we also report $R_\text{eff} = R_m / \sqrt{\sum_{i=1}^r \overline{y}_i^2}$, which is normalized to the approximate radius of the training data. The denominator denotes the root-mean-square of the temporal average of each component of the trajectory.

\begin{table}[h]
\centering
\begin{tabular}{ |>{\columncolor[gray]{0.85}}p{2.07cm}|p{0.35cm}|p{1.45cm}|p{1cm}|p{0.4cm}|p{0.5cm}|p{0.5cm}|p{2cm}|p{0.5cm}|p{0.9cm}|p{0.8cm}|p{0.9cm}|  }
 \hline
 \rowcolor{gray!30} & r & $\Delta t$ & M & $\lambda$ & $\eta$ & $\gamma$ & $\bm{m}^*$ & $R_m$ & $R_\text{eff}$ & $\lambda_1$  & $\text{E}_\text{m} (\%)$ \\
 \hline
  Mean field & 3 & $10^{-2}$ & $50000$ & 0 & $10^{10}$ & 1 & $[0, 0,  1.3]$ & 1.3 & 218 & -1 & $10^{-3}$ \\
 \hline
 Oscillator & 3 & $5\times 10^{-3}$ & $50000$ & 0 & $10^{8}$ & 0.1 & $[0, -0.9, 0.4]$ & 300 & 597 & $-0.01$ & $10^{-2}$\\
 \hline
  Lorenz & 3 & $5\times 10^{-3}$ & $50000$ & 0 & 0.1 & 1 & $[-1.2,0.1,38]$ & $106$ & 4.4 & $-1$ & $0.3$\\
 \hline
  Triadic MHD & 6 & $10^{-3}$ & $50000$ & 0 & $10^3$ & 0.1 & $[0, ...]$ & $-$ & $-$ & $0$ & $10^{-4}$\\
 \hline
  Burgers' Eq. & 10 & 0.1 & 30000 & 0 & 500 & 0.1 & $[-0.2,0,...]$ & $-$ & $-$ & 0.1 & $-$ \\
 \hline
  Von K\`arm\`an & 5 & 0.1 & 30000 & 0.1 & 1 & 0.1 & $[-1.2, ...,  1.1]$ & $29$ & $17$ & $-0.1$ & $-$\\
 \hline
\end{tabular}
\caption{Description of the sampling, trapping SINDy hyperparameters, and identified trapping region for the dynamic systems examined in Section~\ref{sec:results}. Trajectory data does not include any added noise so $\lambda = 0$ works for most of the systems. The SINDy models are identified from a single trajectory. These parameters produce reasonable results for these systems, but a hyperparameter scan can lead to further improvements. }
\label{tab:results_summary}
\vspace{-0.15in}
\end{table}

\subsection{Mean field model\label{sec:results_meanfield}}
\begin{figure}[t]
\begin{subfigure}[b]{0.48\textwidth}
\raggedleft
\begin{overpic}[width=0.99\linewidth]{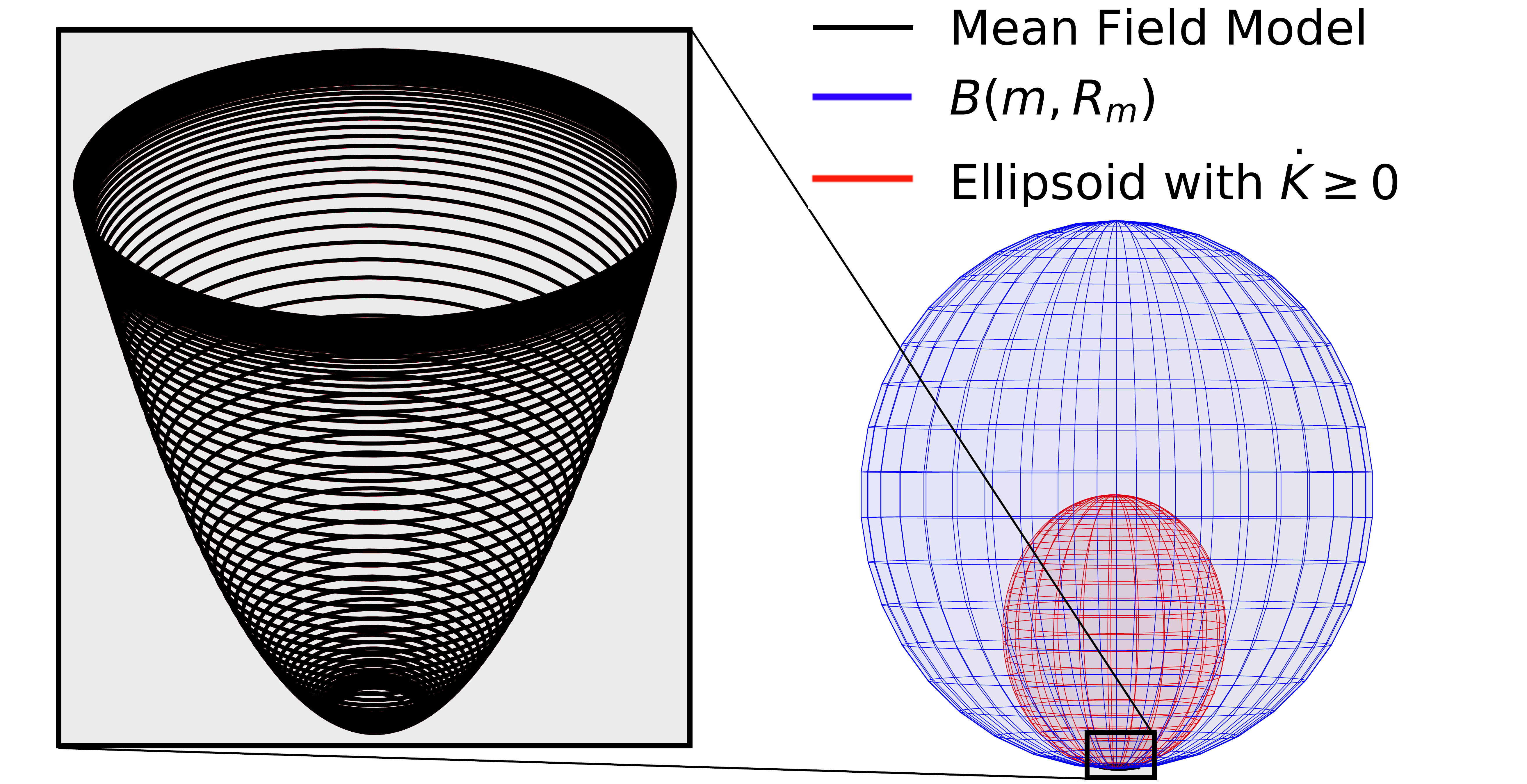}
\end{overpic}
\caption{Trapping SINDy model (black) of a mean field system trajectory (red) with $\mu = 0.01$ and initial condition $[\mu, \mu, 0]$. The trajectory is shown within the estimated trapping region and ellipsoid where $\dot{K} \geq 0$. The prediction error is $\text{E}_\text{pred} \approx 0.6\%$.} 
\label{fig:meanfield_model}
\end{subfigure}
\hspace{0.2in}
\begin{subfigure}[b]{0.48\textwidth}
\raggedright
\begin{overpic}[width=0.99\linewidth]{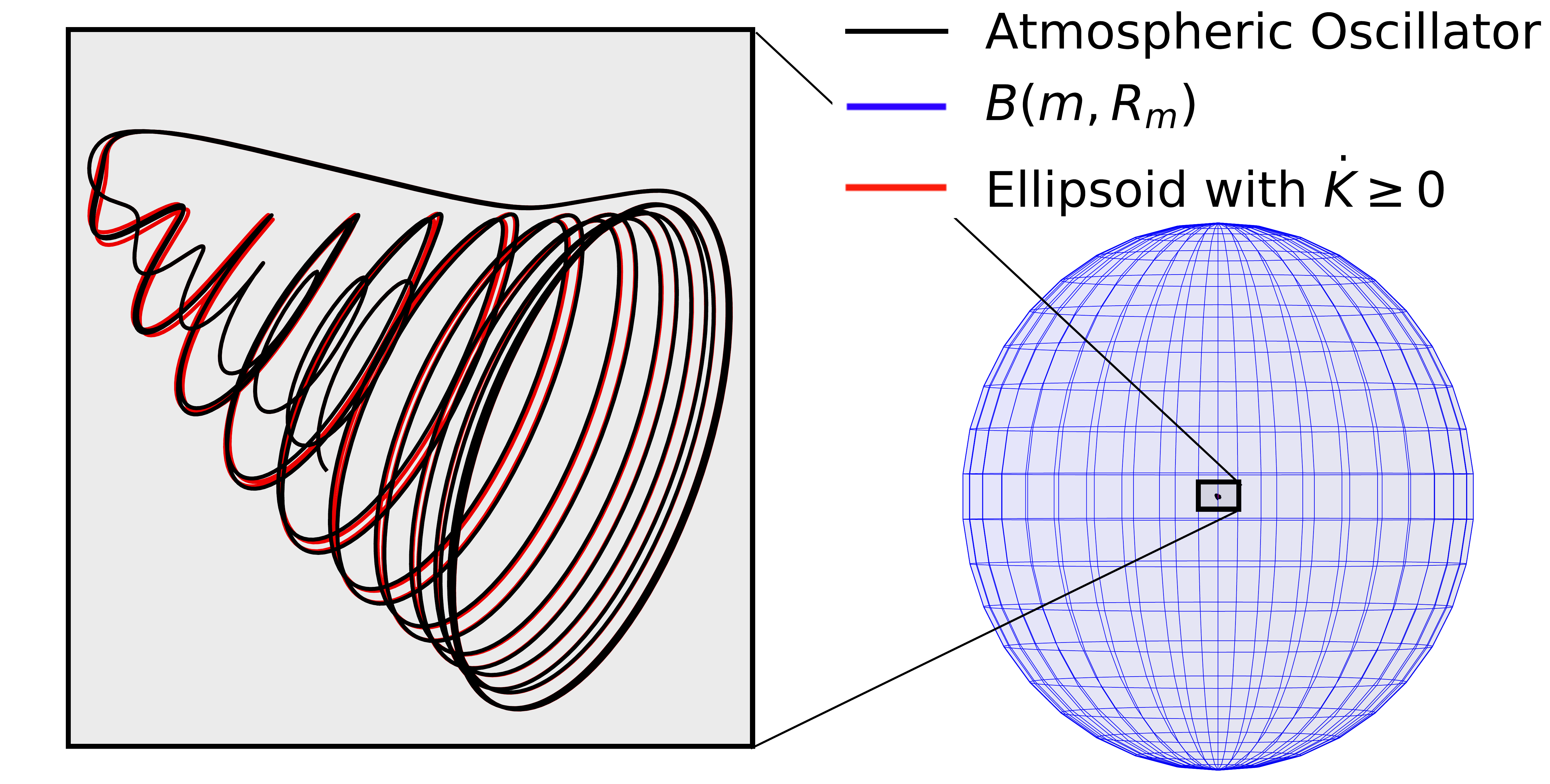}
\end{overpic}
\caption{Same illustration for the atmospheric oscillator with random initial condition chosen from the unit ball. There is large scale separation in this system, so that $|\lambda_1| \ll |\lambda_2|, |\lambda_3|$. This leads to an overestimate of the trapping region size. The prediction error is $\text{E}_\text{pred} \approx 6\%$.} 
\label{fig:oscillator_model}
\end{subfigure}
\caption{Identified models and trapping regions for the mean field and atmospheric oscillator systems.}
\vspace{-0.15in}
\end{figure}

Often the trajectories of a nonlinear dynamical system, whose linear part has some stable directions, will approach a slow manifold of reduced dimension with respect to the full state space.
As an example of this behavior, consider the following linear-quadratic system originally proposed by Noack et al.~\cite{noack2003hierarchy} as a simplified model of the von Karman vortex shedding problem explored further in Sec.~\ref{sec:results_vonKarman}:
\begin{align}
\label{eq:mean-field}
    \frac{d}{dt}\begin{bmatrix}
    x \\ 
    y \\
    z \\
    \end{bmatrix} = \begin{bmatrix}
    \mu & -1 & 0 \\
    1 & \mu & 0 \\
    0 & 0 & -1 \\
    \end{bmatrix}\begin{bmatrix}
    x \\ 
    y \\
    z 
    \end{bmatrix}
    +
    \begin{bmatrix}
    - xz \\ - yz \\ x^2 + y^2
    \end{bmatrix}.
\end{align}
Systems of this form commonly arise in PDEs with a pair of unstable eigenmodes represented by $x$ and $y$.
The third variable $z$ models the effects of mean-field deformations due to nonlinear self-interactions of the instability modes.
The system undergoes a supercritical Hopf bifurcation at $\mu = 0$; for $\mu \ll 1$ trajectories quickly approach the parabolic manifold defined by ${z = x^2 + y^2}$.
All solutions asymptotically approach a stable limit cycle on which $z = x^2 + y^2 = \mu$.
It is enough to notice that $\bm{m} = [0, 0, \mu+\epsilon]$, $\epsilon > 0$ produces
\begin{align}
    \bm{A}^S = \bm{L}^S - \bm{m}^T\bm{Q} = \begin{bmatrix}
    -\epsilon & 0 & 0 \\
    0 & -\epsilon & 0 \\
    0 & 0 & -1 
    \end{bmatrix},
\end{align}
so this system exhibits a trapping region. We illustrate a stable and accurate model identified by our trapping SINDy algorithm in Fig.~\ref{fig:meanfield_model}.

This system is of particular interest because it is a prototypical example of how quadratic interactions in a multi-scale system can give rise to effective higher-order nonlinearities.
If the dynamics are restricted to the slow manifold, the system reduces to the cubic Hopf normal form~\cite{noack2003hierarchy,guckenheimer_holmes}
\begin{align}
\label{eq:mean-field-slow}
    \frac{d}{dt}\begin{bmatrix}
    x \\ 
    y \\
    \end{bmatrix} = \begin{bmatrix}
    \mu - (x^2 + y^2) & -1 \\
    1 & \mu - (x^2 + y^2) \\
    \end{bmatrix}\begin{bmatrix}
    x \\ 
    y \\
    \end{bmatrix}.
\end{align}
Systems of this type arise in weakly nonlinear pattern-forming systems and are often called Stuart-Landau equations. 
In this case, the nonlinear interactions are no longer energy-preserving, since the manifold restriction removes the fast, dissipative degree of freedom.
We might intuitively expect that this type of manifold reduction would inherit the trapping properties of the underlying system, but to our knowledge a general theory of such situations has not yet been worked out, even for the quadratic energy-preserving case.

\subsection{Atmospheric oscillator model\label{sec:results_oscillator}}

Here we examine a more complicated Lorenz-like system of coupled oscillators that is motivated from atmospheric dynamics:
\begin{align}
\label{eq:oscillator}
    \frac{d}{dt}\begin{bmatrix}
    x \\ 
    y \\
    z \\
    \end{bmatrix} = \begin{bmatrix}
    \mu_1 & 0 & 0 \\
    0 & \mu_2 & \omega \\
    0 & -\omega & \mu_2 
    \end{bmatrix}\begin{bmatrix} 
    x \\ 
    y \\
    z 
    \end{bmatrix}
    +
    \begin{bmatrix}
    \sigma xy \\
    \kappa yz + \beta z^2 - \sigma x^2 \\
    - \kappa y^2 - \beta yz
    \end{bmatrix}.
\end{align}
For comparison, we use the parameters in Tuwankotta et al.~\cite{tuwankotta2006chaos}, $\mu_1 = 0.05$, $\mu_2 = -0.01$, $\omega = 3$, $\sigma = 1.1$, $\kappa = -2$, and $\beta = -6$, for which a limit cycle is known to exist. The trapping SINDy algorithm finds $\bm{m}$ such that $\bm{A}^S$ is negative definite for a wide range of parameter and hyperparameter choices, and accurate model results are illustrated in Fig.~\ref{fig:oscillator_model} alongside the mean-field model results. 

So far, we have illustrated that the trapping algorithm successfully produces accurate and provably stable models on simple systems that exhibit well-behaved attractors. In the next sections, we investigate progressively noisier (Section~\ref{sec:results_Lorenz}) and higher-dimensional (Sections~\ref{sec:results_mhd}--\ref{sec:results_vonKarman}) systems that typically provide significant challenges for model discovery algorithms. 

\subsection{Noisy Lorenz attractor}\label{sec:results_Lorenz}
The Lorenz 1963 system~\cite{lorenz1963deterministic} is among the simplest systems exhibiting chaotic dynamics, developed to model thermal convection in the atmosphere based on computer simulations from his graduate students Ellen Fetter and Margaret Hamilton:
\begin{align}
    \frac{d}{dt}\begin{bmatrix}
    x \\ 
    y \\
    z \\
    \end{bmatrix} &= \begin{bmatrix}
    -\sigma & \sigma & 0 \\
    \rho & -1 & 0 \\
    0 & 0 & -\beta
    \end{bmatrix}
    \begin{bmatrix}
    x \\
    y \\
    z
    \end{bmatrix}
    +
    \begin{bmatrix}
    0 \\
    -xz \\
    xy
    \end{bmatrix}.
\end{align}
For this system, it is possible to write $\bm{A}^S$ explicitly as
\begin{align}
    \bm{A}^S = \begin{bmatrix}
    -\sigma & \frac{1}{2}(\rho+\sigma - m_3) & \frac{1}{2}m_2 \\
    \frac{1}{2}(\rho+\sigma - m_3) & -1 & 0 \\
    \frac{1}{2}m_2 & 0 & -\beta 
    \end{bmatrix}.
\end{align}
For Lorenz's choice of parameters, $\sigma = 10$, $\rho = 28$, $\beta  = 8/3$, this system is known to exhibit a stable attractor. For $\bm{m} = [0,m_2,\rho+\sigma]$ ($m_1$ does not contribute to $\bm{A}^S$ so we set it to zero),
\begin{align}
    \bm{A}^S &= \begin{bmatrix}
    -\sigma & 0 & \frac{1}{2}m_2 \\
    0 & -1 & 0 \\
    \frac{1}{2}m_2 & 0 & -\beta 
    \end{bmatrix}, \qquad
    \lambda_1 = -1, \qquad \lambda_{\pm} = -\frac{1}{2}\left[\beta+\sigma \mp \sqrt{m_2^2 + (\beta-\sigma)^2}\right],
\end{align}
so that if $\lambda_{\pm} < 0$, then $-2\sqrt{\sigma\beta} < m_2 < 2\sqrt{\sigma\beta}$. 
Our algorithm successfully identifies the optimal $\bm{m}$, and identifies the inequality bounds on $m_2$ for stability. 
As this analysis is invariant to $m_1$, in principle the trapping region is given by a cylinder, extruded in the $m_1$ direction, rather than a sphere. 

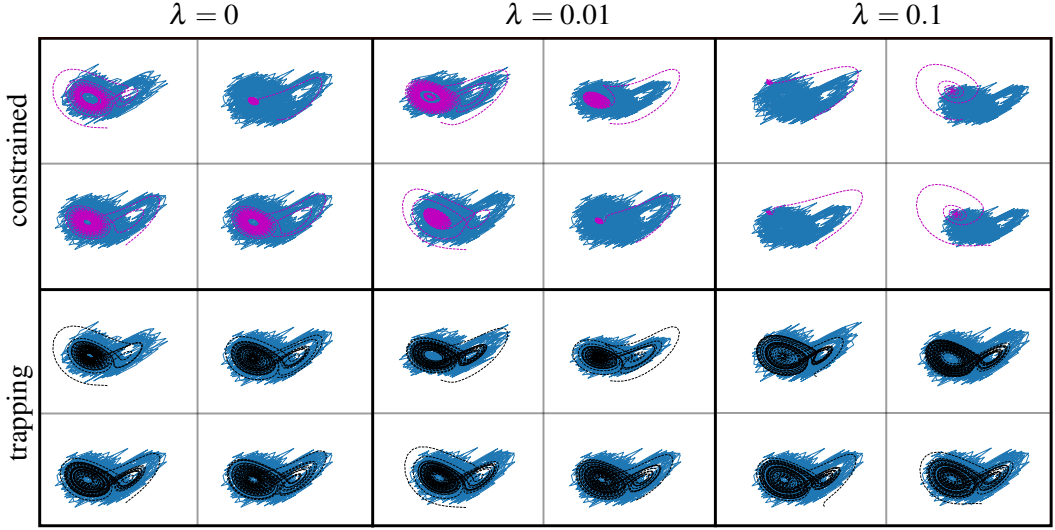
\begin{figure*}[]
\centering
\begin{overpic}[width=0.82\linewidth]{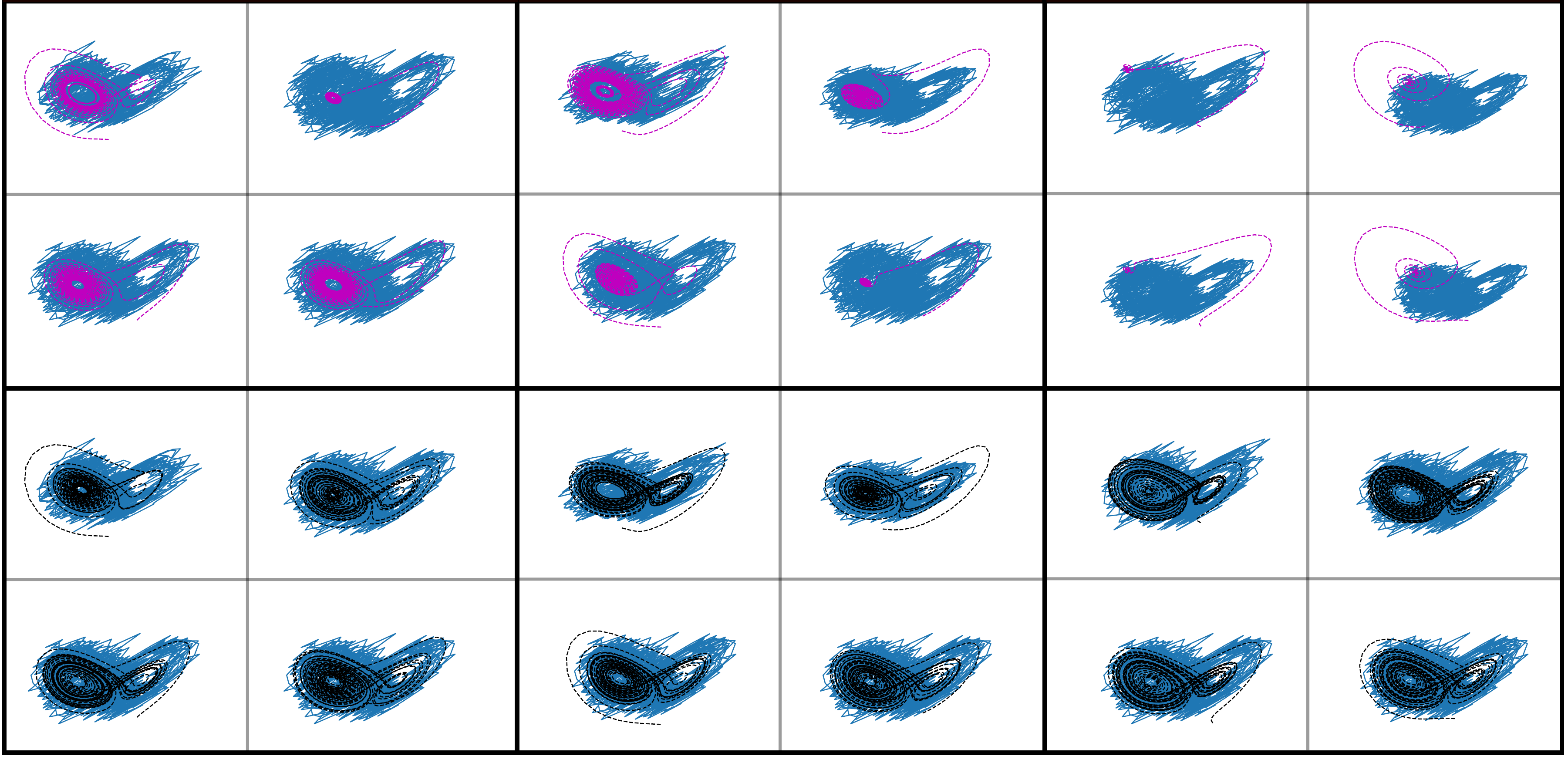}
\put(13,50){$\lambda = 0$}
\put(46,50){$\lambda = 0.01$}
\put(80,50){$\lambda = 0.1$}
\put(-1, 30){\begin{rotate}{90}constrained\end{rotate}}
\put(-1, 7){\begin{rotate}{90}trapping\end{rotate}}
\end{overpic}
\caption{Comparison between the constrained SINDy (magenta) and trapping SINDy (black) results for the Lorenz system using three different values of the sparsity-promotion strength $\lambda$. Unconstrained SINDy results are not pictured because most of the models diverge. Each model is trained on a single Lorenz attractor with noise sampled from $\mathcal{N}(0, 4)$ and an initial condition of $[1,-1,20]$ (blue). The illustrations depict the model performance on data evolved from four random initial conditions between $[-10,10]$ (this testing data is not shown but the attracting set is unchanged). Trapping SINDy produces stable models that follow the underlying attractor for all values of $\lambda$.}
\label{fig:lorenz_comparison}
\vspace{-0.1in}
\end{figure*}

We can show further improvements in model quality. We train unconstrained, constrained, and trapping SINDy models four times; the data for each is a single Lorenz attractor with four different noise instantiations. 
Then we test the performance of the resulting models with a random initial condition in $[-10,10]\times[-10, 10] \times[-10, 10]$. 
For direct comparison, we use the $L^1$ regularizer for each method. Fig.~\ref{fig:lorenz_comparison} illustrates the increased performance with our trapping SINDy algorithm over the constrained SINDy algorithm on noisy Lorenz data for varying threshold levels $\lambda = \{0$, $0.01$, $0.1\}$. The unconstrained method is not pictured because most of the identified models diverge at these high noise levels. At all values of $\lambda$ and most initial conditions, the unconstrained method overfits to the data and produces unstable and diverging models. The traditional constrained SINDy variant mostly manages to produce stable models but produces increasingly poor data fits as $\lambda$ increases. 
In contrast, the trapping version continues to produce stable models that lie on the correct attractor. 
In this way, the additional optimization loss terms that promote stable models provide both a trapping region of known size and additional robustness to noise, even when the models appear otherwise stable, as with many of the constrained SINDy models that incorrectly decay to a fixed point.

\subsection{Triadic MHD model\label{sec:results_mhd}}
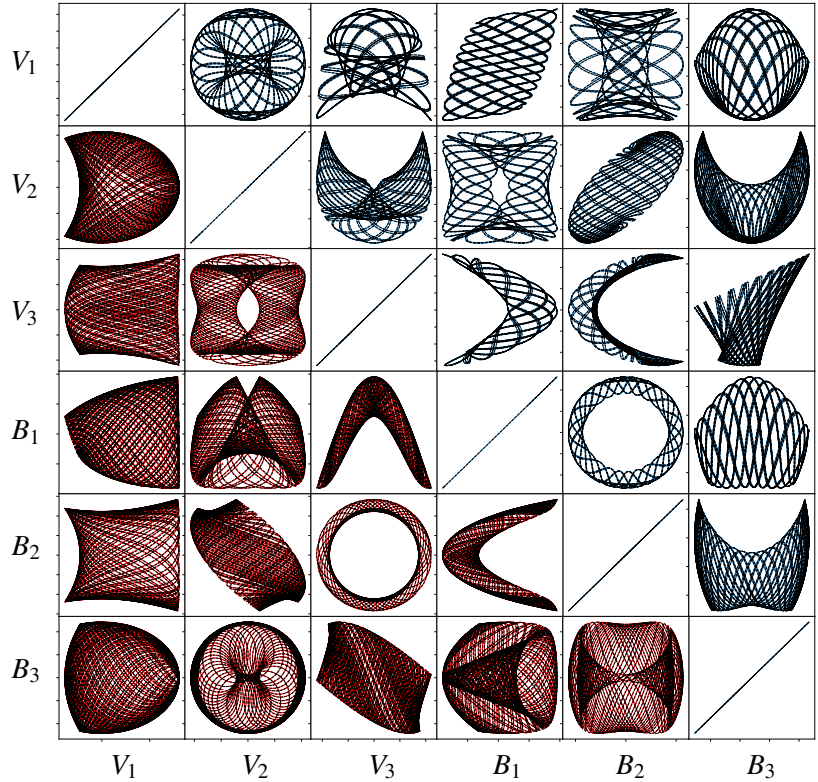
\begin{figure}
    \centering
    \begin{overpic}[width=0.62\linewidth]{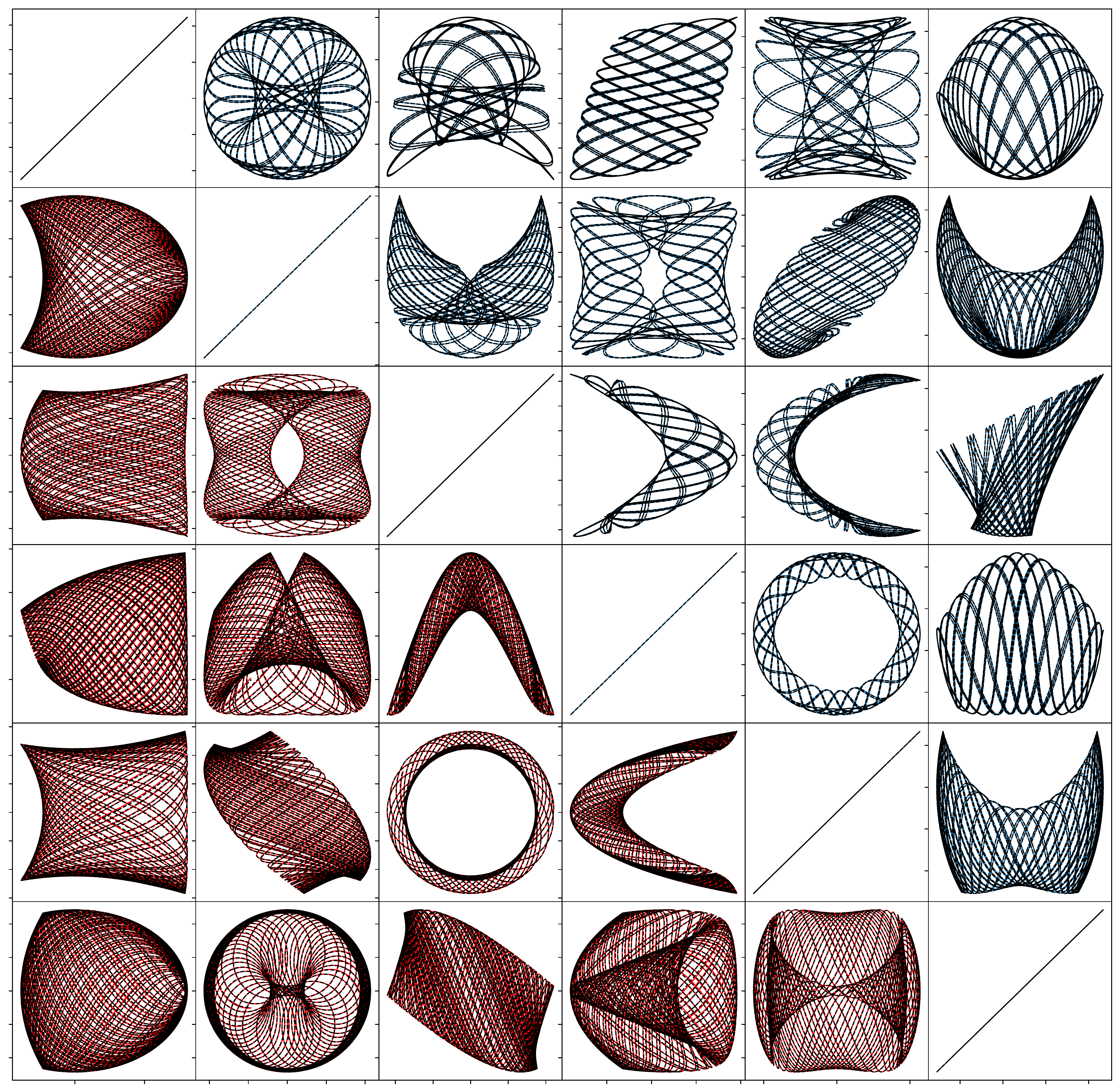}
    \put(-5, 88){$V_1$}
    \put(-5, 72){$V_2$}
    \put(-5, 55.5){$V_3$}
    \put(-5, 40){$B_1$}
    \put(-5, 25){$B_2$}
    \put(-5, 9){$B_3$}
    \put(8, -3.5){$V_1$}
    \put(25, -3.5){$V_2$}
    \put(41.5, -3.5){$V_3$}
    \put(57.5, -3.5){$B_1$}
    \put(73.5, -3.5){$B_2$}
    \put(90.5, -3.5){$B_3$}
    \end{overpic}
    \vspace{0.1in}
    \caption{The triad model for 2D inviscid MHD training data (blue, upper triangle) and a trapping SINDy model (black) capturing Hamiltonian dynamics on testing data (red, lower triangle).}
    \label{fig:mhd_model}
    \vspace{-0.1in}
\end{figure}

Magnetohydrodynamic systems exhibit quadratic nonlinearities that are often energy-preserving with typical boundary conditions. 
We consider a simple model of the nonlinearity in two-dimensional incompressible MHD, which can be obtained from Fourier-Galerkin projection of the governing equations onto a single triad of wave vectors. For the Fourier wave vectors $\bm{k}_1 = (1,1)$, $\bm{k}_2 = (2,-1)$, and $\bm{k}_3 = (3,0)$ and no background magnetic field, the Carbone and Veltri~\cite{carbone1992relaxation} system is 
\setlength{\arraycolsep}{-1pt}
\begin{align}
\label{eq:simpleMHD_model}
\frac{d}{dt}    \begin{bmatrix}
    {V}_1 \\
    {V}_2 \\
    {V}_3 \\ 
    {B}_1 \\
    {B}_2 \\
    {B}_3 \\
    \end{bmatrix} = \begin{bmatrix}
    -2 \nu & 0 & 0 & 0 & 0 & 0 \\
    0 & -5 \nu & 0 & 0 & 0 & 0 \\
    0 & 0 & -9 \nu & 0 & 0 & 0 \\
    0 & 0 & 0 & -2 \mu & 0 & 0 \\
    0 & 0 & 0 & 0 & -5 \mu & 0 \\
    0 & 0 & 0 & 0 & 0 & -9 \mu \\
    \end{bmatrix}\begin{bmatrix}
    V_1 \\ 
    V_2 \\ 
    V_3 \\ 
    B_1 \\ 
    B_2 \\ 
    B_3 
    \end{bmatrix} + \begin{bmatrix}
    4(V_2V_3 - B_2B_3) \\ 
    -7(V_1V_3 - B_1B_3) \\ 
    3(V_1V_2 - B_1B_2) \\ 
    2(B_3V_2 - V_3B_2) \\ 
    5(V_3B_1 - B_3V_1) \\ 
    9(V_1B_2 - B_1V_2) \\ 
    \end{bmatrix},
\end{align}
where $\nu \geq 0$ is the viscosity and $\mu \geq 0$ is the resistivity. 
Without external forcing, this system is stable, dissipating to zero, so we consider the inviscid limit $\nu = \mu = 0$. The system is now Hamiltonian and our algorithm correctly converges to $\bm{m} = 0$, $\bm{A}^S = 0$. 
The results in Fig.~\ref{fig:mhd_model} provide a useful illustration that trapping SINDy converges to stable energy-preserving models even when the trapping theorem is not satisfied. 
These results also provide a reminder that there are a large number of dynamical systems beyond fluids, such as MHD, which may benefit from these types of techniques. 
The reason our algorithm converges to the correct behavior is because it is still minimizing $\dot{K}$; in this case trapping SINDy converges to $\dot{K} \approx 0$ and can make no further improvement.

\subsection{Forced Burgers' equation\label{sec:results_burgers}}
The viscous Burgers' equation has long served as a simplified one-dimensional analogue to the Navier-Stokes equations~\cite{Burgers1948, Hopf1948}. The forced, viscous Burgers' equation on a periodic domain $x \in [0,2\pi)$ is:
\begin{align}
\label{eq:burgers}
    \frac{d}{dt}{u} &= -(U + u)\partial_x u + \nu \partial_{xx}^2u + g(x,t),
\end{align}
where $\nu$ is viscosity and the constant $U$ models mean-flow advection. 
We project this system onto a Fourier basis and assume constant forcing acting on the largest scale, i.e., $g(x, t) = \sigma \left( a_1(t) e^{ix} + a_{-1}(t) e^{-ix} \right)$, as in Noack et al.~\cite{noack2008finite}.
After Fourier projection, the evolution of the coefficients $a_k(t)$ is given by the Galerkin dynamics
\begin{equation}
\label{eq:burgers_galerkin}
    \dot{a}_k = \left( \delta_{|k|1} \sigma - \nu k^2  - ikU \right) a_k - \sum_{\ell=-r}^{r} i \ell a_{\ell} a_{k - \ell}.
\end{equation}

In the subcritical case $\sigma < \nu$, the origin of this system is stable to all perturbations and all solutions decay for long times.
However, in the supercritical case $\sigma > \nu$, the excess energy input from the forcing cascades to the smaller dissipative scales. 
The ``absolute equilibrium'' limit $\sigma = \nu = 0$ has a Hamiltonian structure; for long times the coefficients approach thermodynamic equilibrium and equipartition of energy~\cite{majda2000remarkable}.
This structure does not correspond to any physical behavior of the Navier-Stokes equations, although it does approximate some properties of the inviscid Euler equations~\cite{Kraichnan1989}.
Due to its rich dynamics, this modified Burgers' equation has also been investigated in the context of closure schemes for Galerkin models~\cite{noack2008finite}.
We simulate the PDE in Eq.~\eqref{eq:burgers} with a high-resolution Godunov-type finite volume method using a van Leer flux limiter, implemented in the open-source Clawpack solver~\cite{clawpack}.

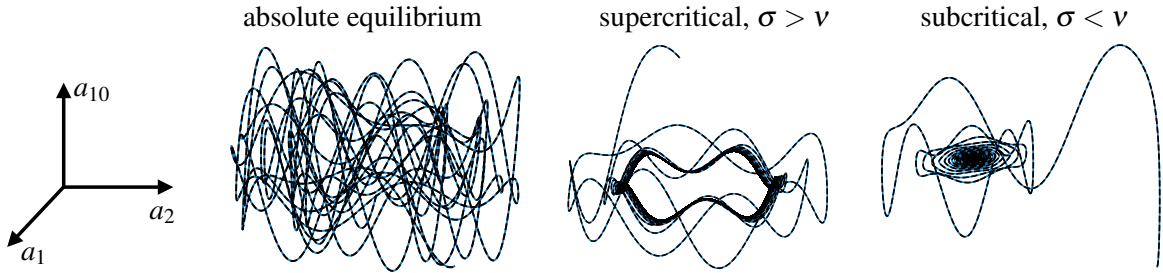
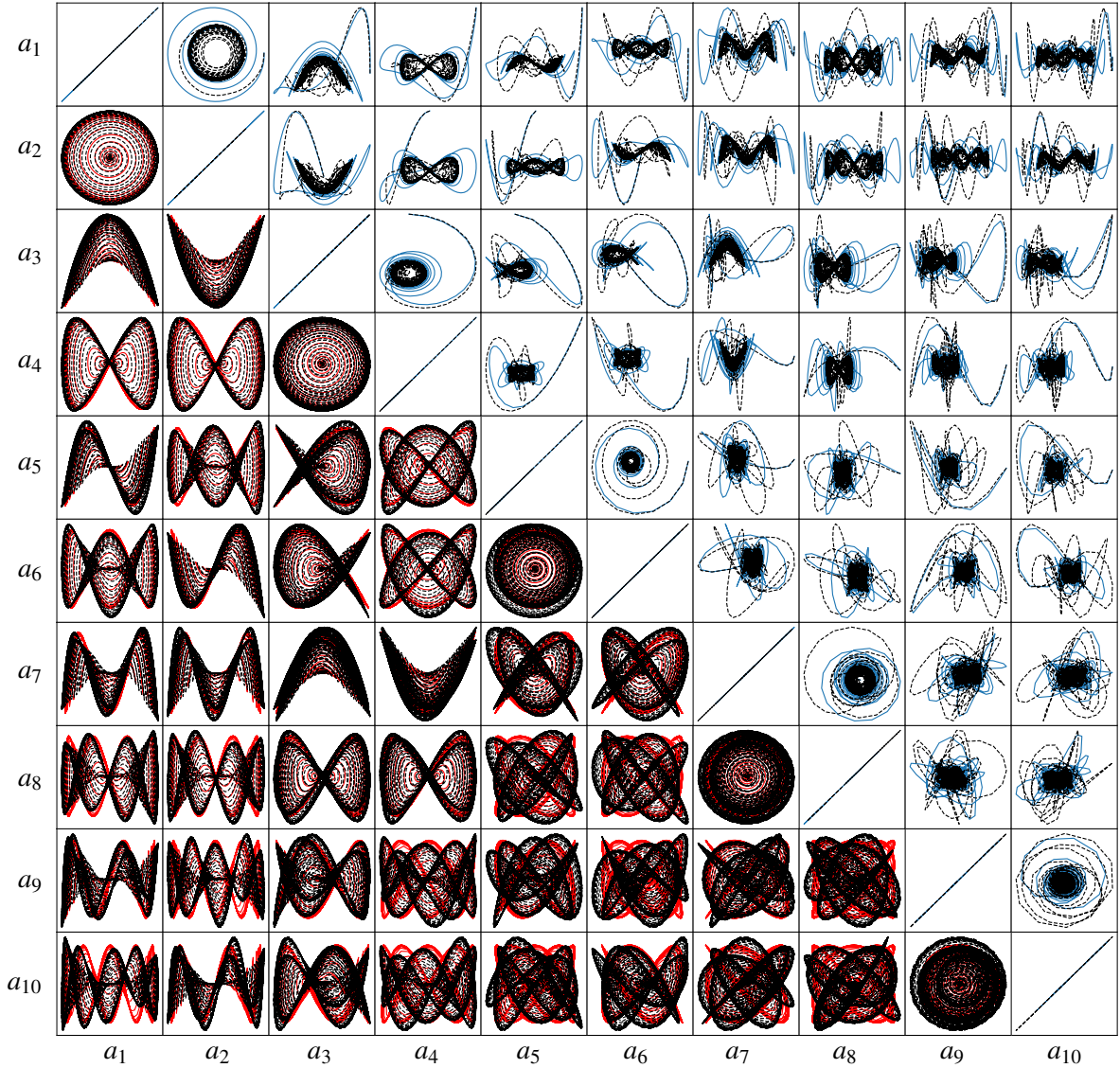
\begin{figure*}[]
\begin{subfigure}[b]{0.99\textwidth}
\centering
\begin{overpic}[width=0.95\linewidth]{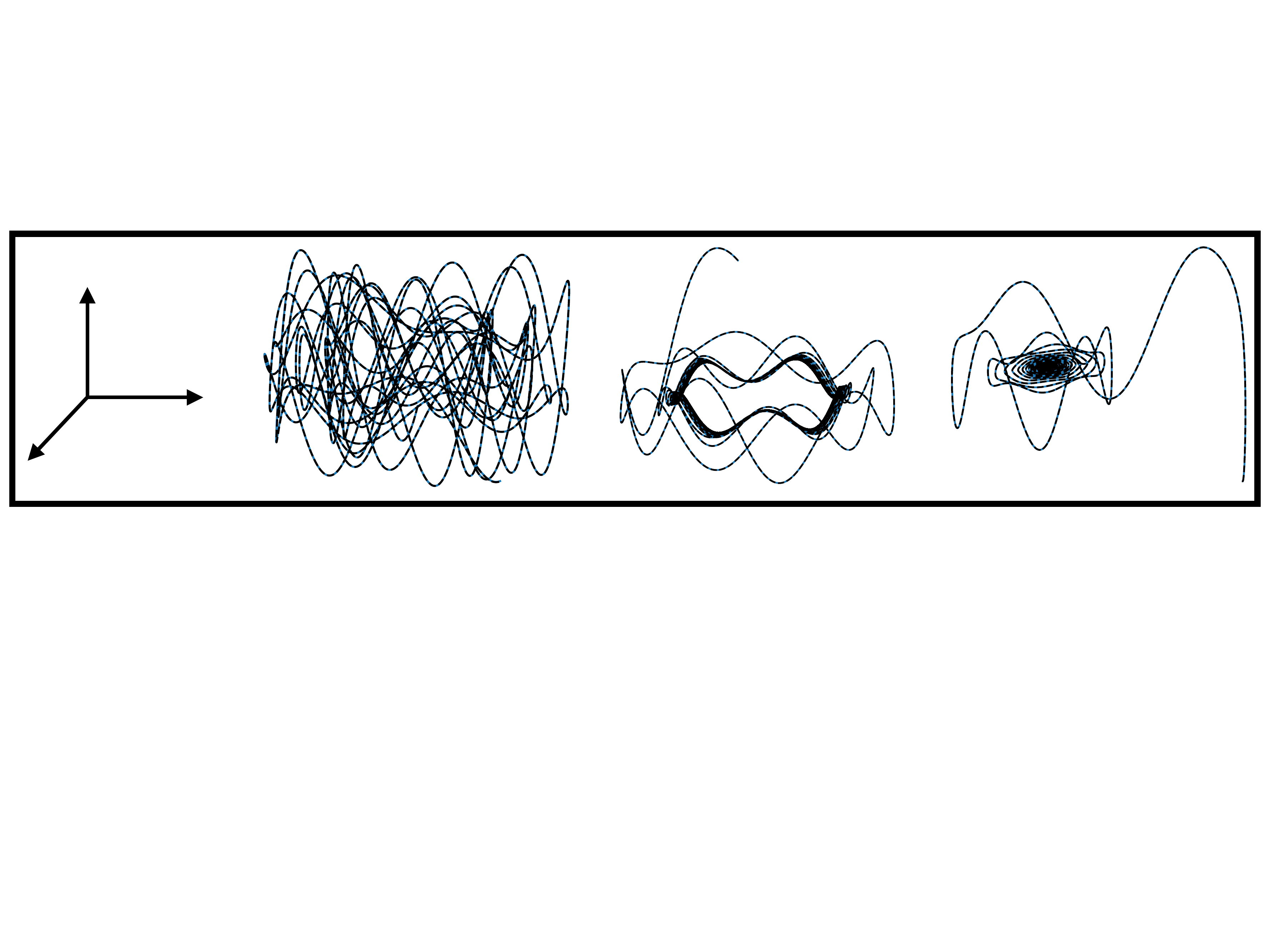}
\put(6.5,16){$a_{10}$}
\put(2,2){$a_{1}$}
\put(13,5.5){$a_{2}$}
\put(21, 22){absolute equilibrium}
\put(51.5, 22){supercritical, $\sigma > \nu$}
\put(79, 22){subcritical, $\sigma < \nu$}
\end{overpic}
\caption{Trapping SINDy model (black) for the modified Burgers' equation in the three dynamic regimes. For improved illustration, the ground truth data (blue) is generated from the 10D Noack et al.~\cite{noack2008finite} model rather than DNS. }
\label{fig:burger_regimes}
\vspace{0.15in}
\end{subfigure}
\begin{subfigure}[b]{0.96\textwidth}
\hspace{.225in}\begin{overpic}[width=0.95\linewidth]{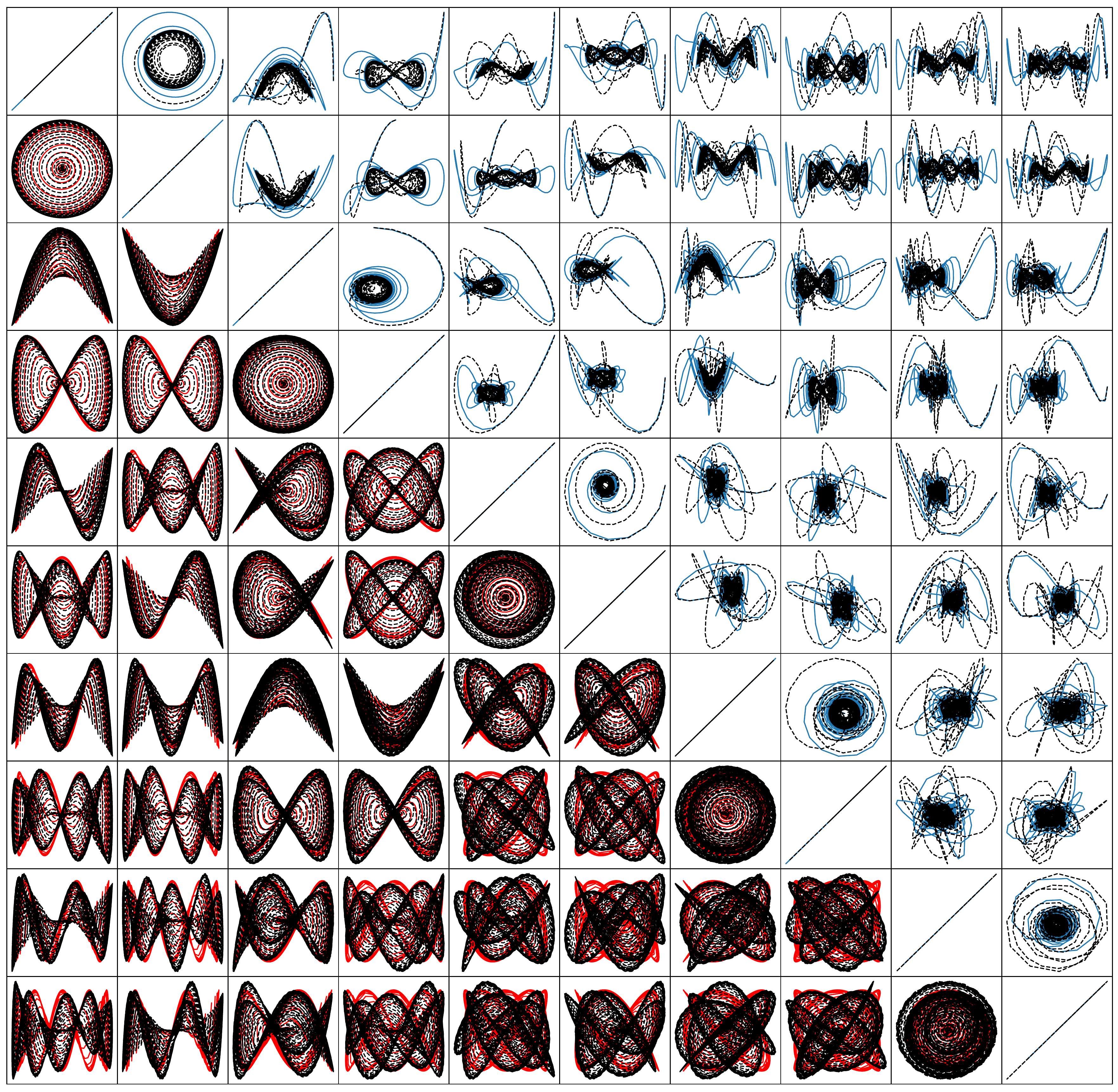}
\normalsize
\put(-3, 92.5){$a_1$}
\put(-3, 83){$a_2$}
\put(-3, 73){$a_3$}
\put(-3, 63){$a_4$}
\put(-3, 53.5){$a_5$}
\put(-3, 43.5){$a_6$}
\put(-3, 34){$a_7$}
\put(-3, 24){$a_8$}
\put(-3, 14.5){$a_9$}
\put(-4, 5){$a_{10}$}
\put(5, -1.5){$a_1$}
\put(14.5, -1.5){$a_2$}
\put(24, -1.5){$a_3$}
\put(34, -1.5){$a_4$}
\put(43.5, -1.5){$a_5$}
\put(53.5, -1.5){$a_6$}
\put(63, -1.5){$a_7$}
\put(73, -1.5){$a_8$}
\put(83, -1.5){$a_9$}
\put(93, -1.5){$a_{10}$}
\end{overpic}
\vspace{0.2in}
\caption{Temporal evolutions of each $(a_i,a_j)$ pair for $i,j=1,...,10$ obtained from DNS training data (blue, upper triangle), DNS testing data (red, lower triangle), and trapping SINDy prediction on both DNS datasets (black). The trapping algorithm struggles a bit with the transients, but obtains the correct attractor behavior.}
\label{fig:burger_DNS}
\end{subfigure}
\caption{Summary of trapping SINDy performance for the forced Burgers' equation.}
\label{fig:burger_results}
\end{figure*}

We illustrate the model performance in Fig.~\ref{fig:burger_regimes} for the subcritical case with $\sigma = 0.01$ and $\nu = 0.025$, the supercritical case with $\sigma = 0.1$ and $\nu = 0.025$, and the absolute equilibrium. 
In all cases $U = 1$. 
For the subcritical condition, all the eigenvalues of $\bm{L}^S$ are negative, and thus the algorithm finds stable models. For the supercritical condition $\sigma > \nu$, there is some subtlety. The algorithm does not converge to a negative definite $\bm{A}^S$, although it finds a solution with $\dot{K} \leq 0$. As mentioned in Section~\ref{sec:effective_nonlinearity}, this system does not exhibit effective nonlinearity. 
This lack of effective nonlinearity was also true for the MHD example in Section~\ref{sec:results_mhd}, since the initial condition with no magnetic field perturbation, $B_1(0) = B_2(0) = B_3(0) = 0$, remains on the purely hydrodynamic manifold. In the inviscid limit, we did not need to consider this subspace because the system already does not satisfy the trapping theorem by virtue of being Hamiltonian.
Lastly, in the absolute equilibrium regime the trapping SINDy algorithm correctly identifies vanishing eigenvalues of $\bm{A}^S$. In practice, we find excellent models for all of the aforementioned systems and for all practical purposes these models are typically stable, regardless of effective nonlinearity or Hamiltonian dynamics, because the SINDy trapping algorithm is minimizing $\dot{K}$. However, without effective nonlinearity we are not guaranteed to produce a stable model for every possible initial condition. 

In Fig.~\ref{fig:burger_DNS} we illustrate the $r=10$ model built from the DNS data in the supercritical regime with $\sigma = 0.1$, $\nu = 0.025$. It struggles a bit with the transient but otherwise the performance is accurate. Part of the reason for the poor fit to the transient is that $\lambda = 0$ is used here. The biasing towards stability appears to mitigate some of the need for sparsity-promotion; in other words, sparsity-promotion is not necessarily needed to produce a stable model, but may be needed for a more accurate or interpretable model, since the number of coefficients in $Q_{ijk}$ is $\mathcal{O}(r^3)$ despite the constraints. Using finite $\lambda$ may improve the model further, especially the transients, but instead of further investigating this example, we move on and conclude the results by addressing the challenging von K\`arm\`an vortex shedding behind a circular cylinder.

\subsection{Von K\`arm\`an vortex street }
\label{sec:results_vonKarman}
Here we investigate the fluid wake behind a bluff body, characterized by a periodic vortex shedding phenomenon known as a von K\`arm\`an street. 
The two-dimensional incompressible flow past a cylinder is a stereotypical example of such behavior, and has been a benchmark problem for Galerkin models for decades~\cite{noack2003hierarchy}.
The transition from a steady laminar solution to vortex shedding is given by a Hopf bifurcation, as a pair of eigenvalues of the linearized Navier-Stokes operator cross the real axis.

The transient energy growth and saturation amplitude of this instability mode is of particular interest and has historically posed a significant modeling challenge.
Early Galerkin models of vortex shedding, based on a POD expansion about the mean flow, captured the oscillatory behavior but were structurally unstable~\cite{Deane1991pof}.
This was later resolved by Noack et al.~\cite{noack2003hierarchy}, who recognized that the transient behavior could be explained by Stuart-Landau nonlinear stability theory, in which the unsteady symmetric flow is deformed to the neutrally stable mean flow via a nonlinear self-interaction of the instability mode.
In that work, an 8-mode POD basis was augmented with a ninth ``shift mode'' parameterizing this mean flow deformation.
This approach was later formalized with a perturbation analysis of the flow at the threshold of bifurcation~\cite{Sipp2007jfm}.

This modification encodes the intuition that the dynamics take place on the parabolic manifold associated with the Hopf bifurcation; without it, the energy quadratic models tends to overshoot and oscillate before approaching the post-transient limit cycle.
Nevertheless, the 9-mode quadratic Galerkin model does resolve the transient dynamics, nonlinear stability mechanism, and post-transient oscillation, accurately reproducing all of the key physical features of the vortex street. Moreover, in Schlegel and Noack~\cite{Schlegel2015jfm} stability of the quadratic model was proven with $m_9 = m_\text{shift} = \epsilon$, $\epsilon > 1$, and $m_i = 0$ for $i = \{1,...,8\}$. Recall from the discussion in Section~\ref{sec:effective_nonlinearity} that POD-Galerkin models will generally weakly satisfy the effective nonlinearity criteria and it is unclear if the shift-mode complicates this picture.

Although the POD-Galerkin model is an accurate description of the flow past a cylinder, it is an intrusive model, in the sense that evaluating the projected dynamics requires evaluating individual terms in the governing equations, such as spatial gradients of the flow fields.
POD-Galerkin models therefore tend to be highly sensitive to factors including mesh resolution, convergence of the POD modes, and treatment of the pressure and viscous terms.
Recent work by Loiseau et al.~\cite{loiseau2018constrained,loiseau2018sparse,loiseau2019pod} has bypassed the Galerkin projection step by using the SINDy algorithm to directly identify the reduced-order dynamics.
This approach has been shown to yield compact, accurate models for low-dimensional systems ($r=2$ or $3$), but preserving accuracy and stability for higher-dimensional systems remains challenging. 
Higher-dimensional regression problems often become ill conditioned; for example, in the cylinder wake example, the higher modes 3-8 are essentially harmonics of the driving modes 1-2, and so it is difficult to distinguish between the various polynomials of these modes during regression. 
Because these higher harmonics are driven by modes 1-2, the 3D constrained quadratic SINDy model with modes 1-2 plus the shift mode from Loiseau et al.~\cite{loiseau2018constrained} already performs well enough to capture the energy evolution with minor overshoot and correct long-time behavior. Details of the DNS and the POD-Galerkin technique used to reproduce the 9D shift-mode model can be found in Appendix~\ref{appendix:vonKarman_DNS}. 

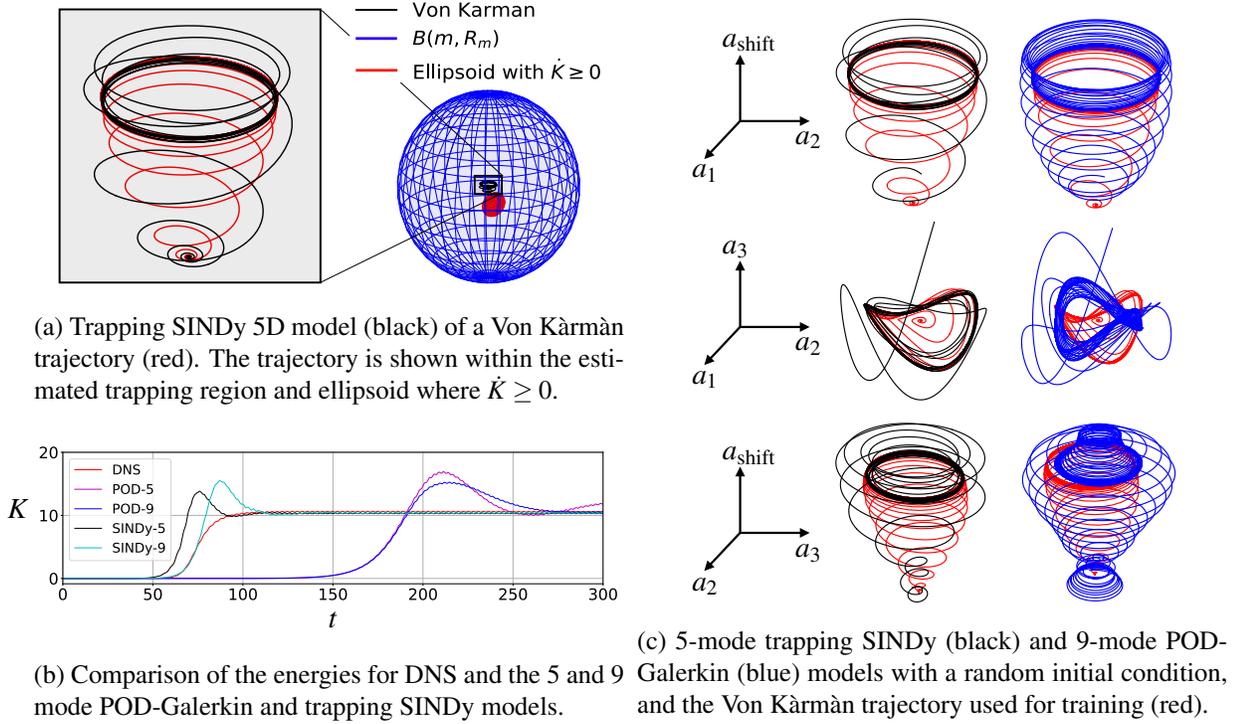
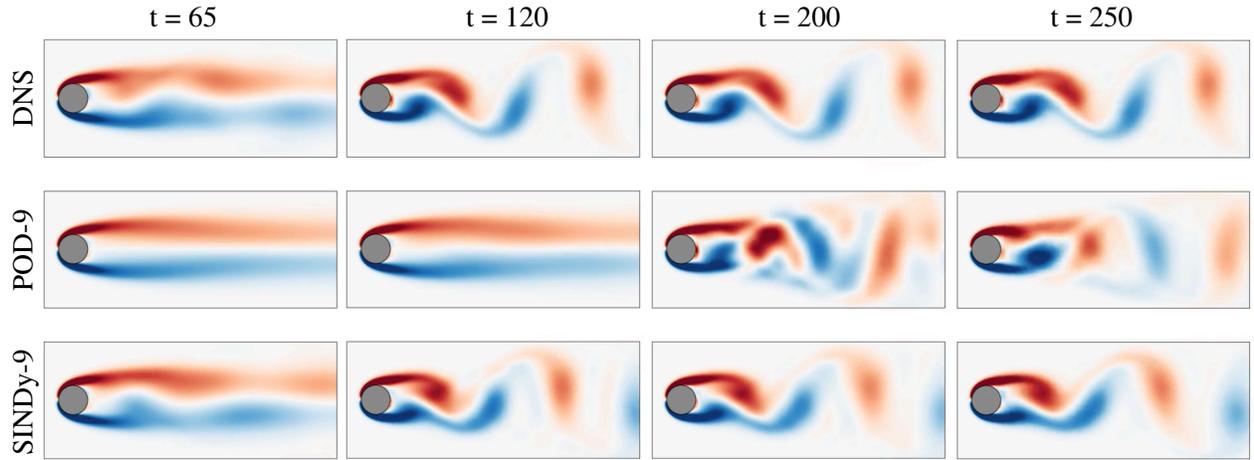
\begin{figure}[h]
\begin{minipage}{0.48\textwidth}
\begin{subfigure}[b]{0.99\textwidth}
\centering
\begin{overpic}[width=1.0\linewidth]{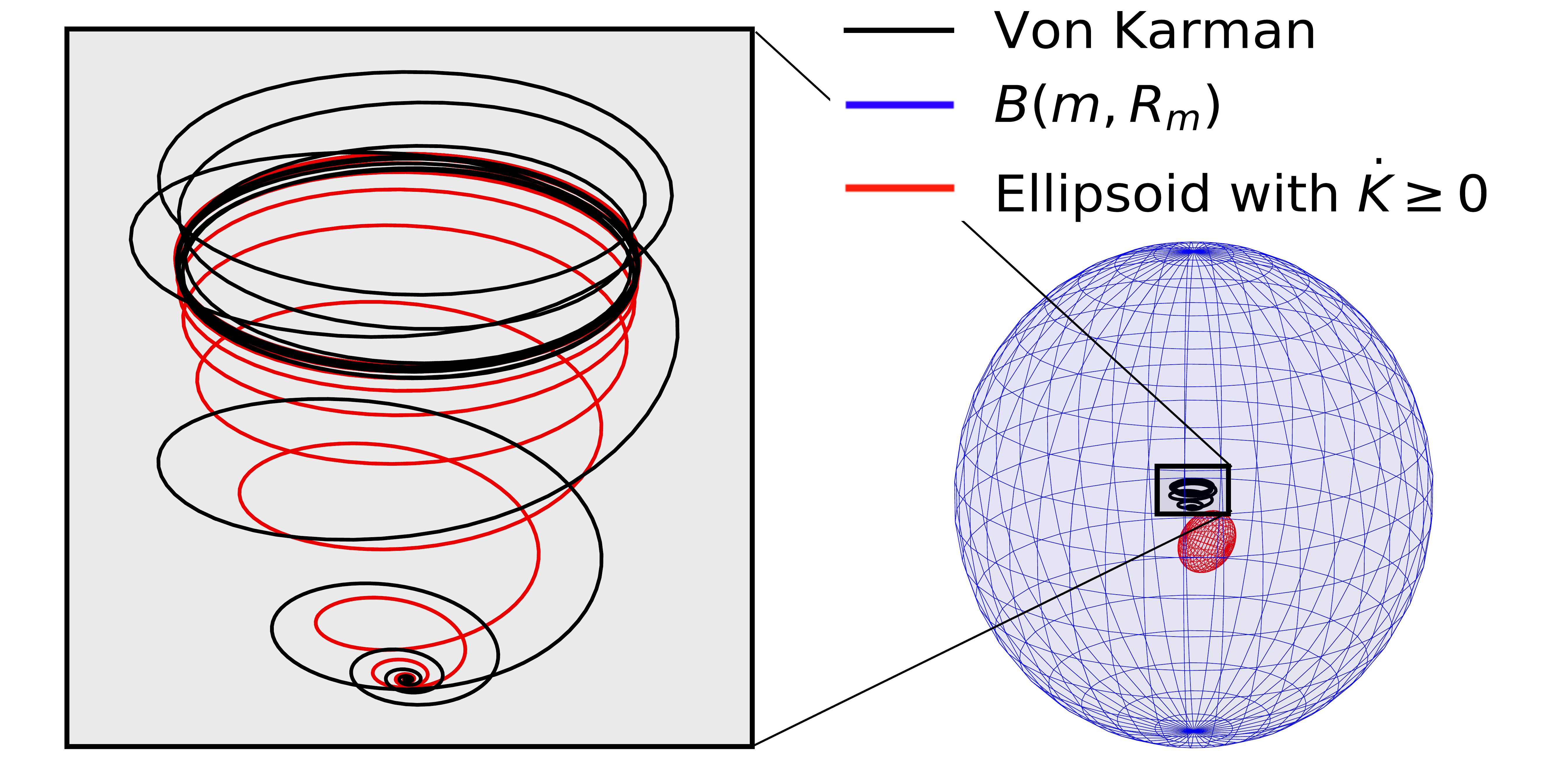}
\end{overpic}
\caption{Trapping SINDy 5D model (black) of a Von K\`arm\`an trajectory (red). The trajectory is shown within the estimated trapping region and ellipsoid where $\dot{K} \geq 0$.}
\label{fig:vonKarman_trappingRegion}
\end{subfigure}
\begin{subfigure}[b]{0.99\textwidth}
\centering
\vspace{0.5cm}
\begin{overpic}[width=0.99\linewidth]{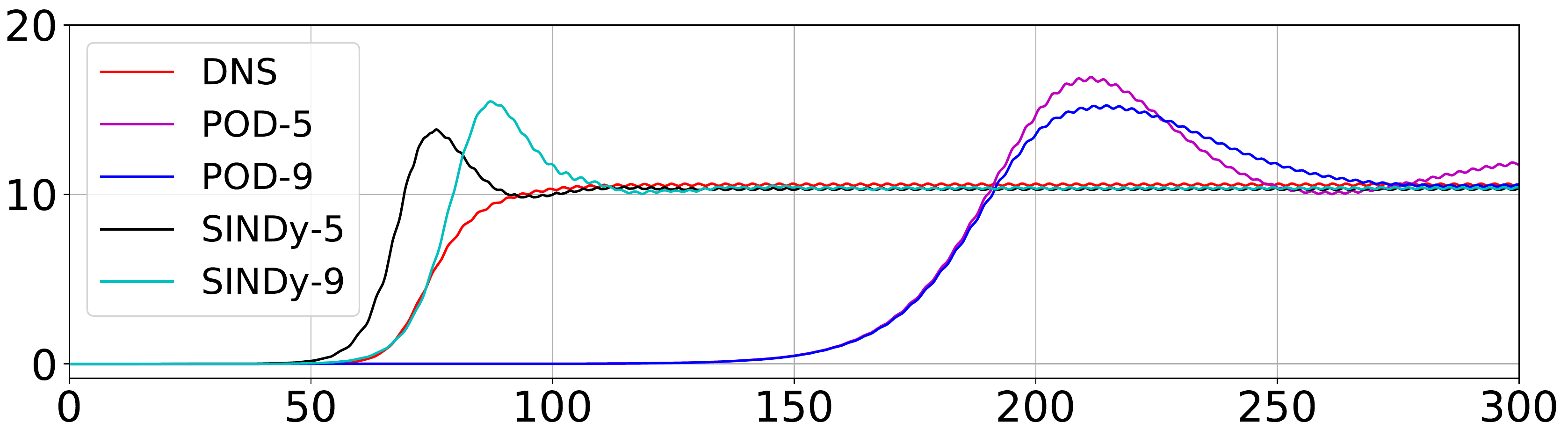}
\put(-5, 15){$K$}
\put(50, -3.75){$t$}
\end{overpic}
\vspace{0.02in}
\caption{Comparison of the energies for DNS and the 5 and 9 mode POD-Galerkin and trapping SINDy models.}
\label{fig:vonKarman_energies}
\end{subfigure}
\end{minipage}
\label{fig:vonKarman_results}
\begin{minipage}{0.48\textwidth}
\begin{subfigure}[b]{0.99\textwidth}
\centering
\begin{overpic}[width=0.82\linewidth]{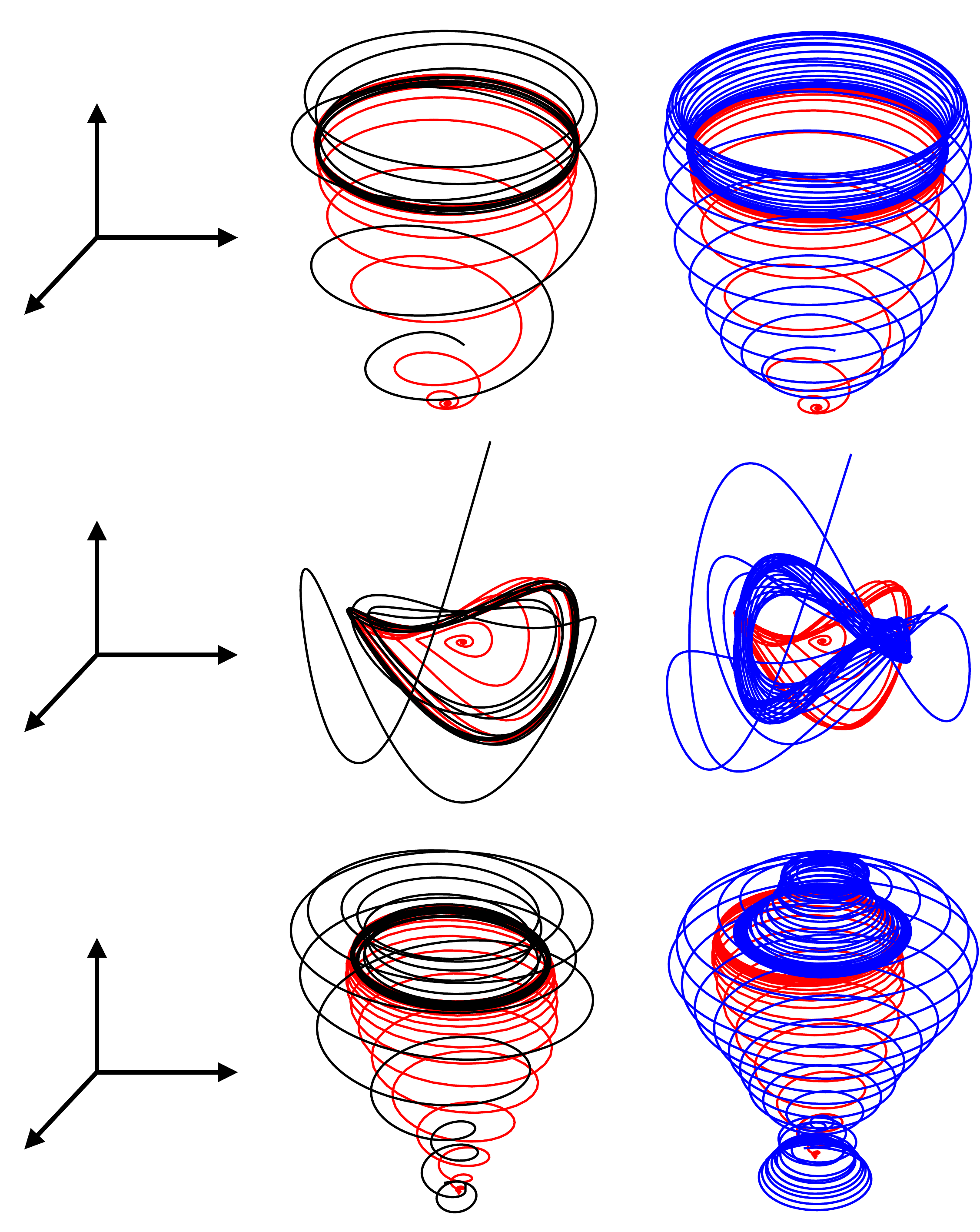}
\put(0, 3.5){$a_2$}
\put(17, 9){$a_3$}
\put(5, 25){$a_\text{shift}$}
\put(0, 37.5){$a_1$}
\put(17, 43){$a_2$}
\put(5, 59){$a_3$}
\put(0, 71.5){$a_1$}
\put(17, 77){$a_2$}
\put(5, 93){$a_\text{shift}$}
\end{overpic}
\caption{5-mode trapping SINDy (black) and 9-mode POD-Galerkin (blue) models with a random initial condition, and the Von K\`arm\`an trajectory used for training (red).}
\label{fig:vonKarman_trajectories}
\end{subfigure}
\end{minipage}

\vspace{0.4in}

\begin{subfigure}[b]{0.99\textwidth}
\centering
\begin{overpic}[width=0.99\linewidth]{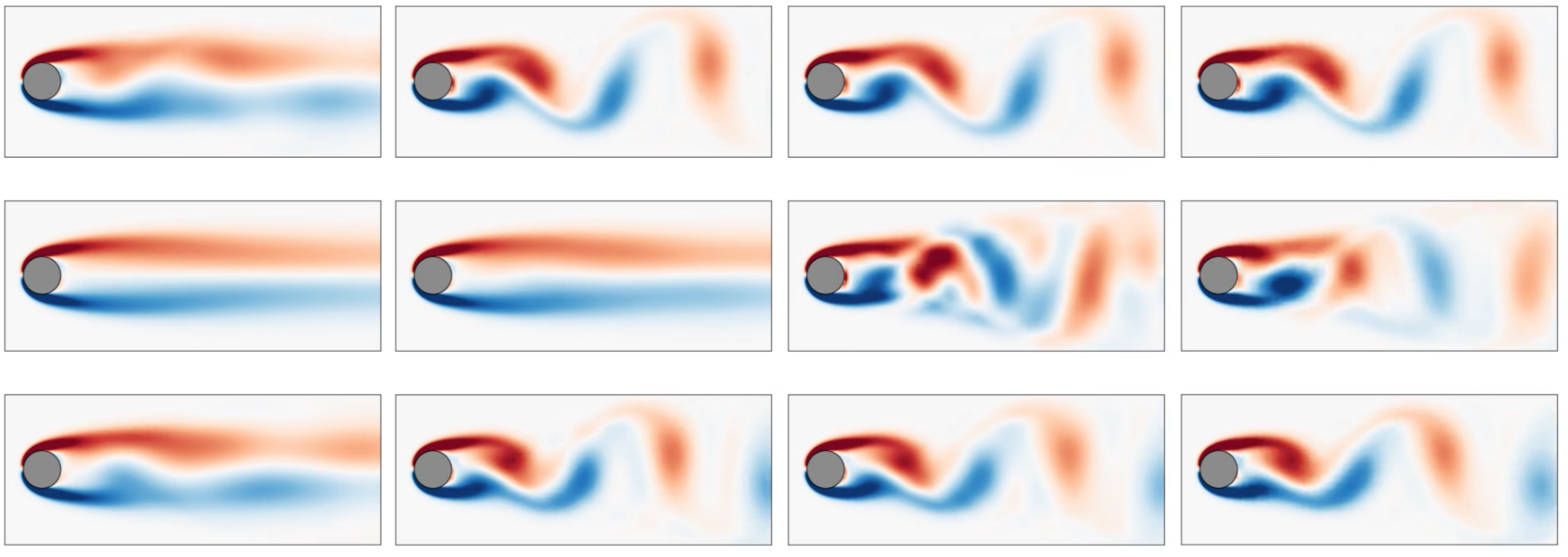}
\put(-2.25, 28){\begin{sideways}DNS\end{sideways}}
\put(-2.25, 14.5){\begin{sideways}POD-9\end{sideways}}
\put(-2.25, 1){\begin{sideways}SINDy-9\end{sideways}}
\put(9, 36.25){t = 65}
\put(35, 36.25){t = 120}
\put(59, 36.25){t = 200}
\put(83, 36.25){t = 250}
\end{overpic}
\caption{Predictions of the vorticity field for the von K\`arm\`an street at four snapshots in time. The trapping SINDy model outperforms the 9D POD-Galerkin model, although an initial phase error in the trapping SINDy prediction (visible in the first snapshot) persists throughout the prediction.}
\label{fig:vonKarman_recons}
\end{subfigure}
\caption{Summary of the differences between DNS, POD-Galerkin models, and trapping SINDy models.}
\end{figure}

With the trapping SINDy algorithm, we obtain new 5-dimensional and 9-dimensional models for the cylinder wake and compare the performance against the same-size analytic POD-Galerkin models. The 5D trapping SINDy model is provably stable and we illustrate the identified trapping region in Fig.~\ref{fig:vonKarman_trappingRegion}. We also compare the 5D SINDy and 9D POD-Galerkin models in Fig.~\ref{fig:vonKarman_trajectories}. The 5D trapping SINDy model outperforms the 9D POD-Galerkin model by significantly improving the transient and improving the identification of the long-term attractor. For the 9D trapping SINDy model, we managed to reduce the largest eigenvalue of $\bm{A}^S$ to $\mathcal{O}(10^{-2}-10^{-4})$ but were unable to produce accurate trapping SINDy models with fully negative definite $\bm{A}^S$. In practice, these models are functionally stable; we tested a large set of random initial conditions and did not find unbounded trajectories. Further searching in the hyperparameter space, or more algorithm iterations for better convergence, could potentially produce fully stable models. 

Despite this setback, the 9D trapping SINDy model performs quite well. 
The Galerkin model and the trapping SINDy model exhibit comparable performance and the SINDy model improves the transient prediction. The energies in Fig.~\ref{fig:vonKarman_energies} illustrate convergence to the true fluid flow energy for all the SINDy and POD-Galerkin models, with only the 9D trapping SINDy model capturing the precise timing of the transient. The flow reconstructions in Fig.~\ref{fig:vonKarman_recons} are quite accurate for both models. This is surprisingly strong performance with SINDy; recall that: 1) the Galerkin model is far more invasive a procedure than SINDy, requiring computation of spatial derivatives and inner products from the DNS, 2) the Galerkin model can still be quite sensitive to the DNS data, boundary conditions, and mesh size, and 3) the 9D trapping SINDy model is far sparser and has far fewer ``active'' terms than the 9D POD-Galerkin model. 

The difficulty in producing provably stable, 9D trapping SINDy models here appears to reveal an interesting optimization tradeoff. While sparsity-promotion tends to promote more accurate models and reduce the complexity of the nonconvex optimization problem (since there are fewer active terms to manage), it also deemphasizes our proposed metric for the strength of effective nonlinearity, $S_e$ from Eq.~\eqref{eq:effective_nonlinearity_strength}, by reducing the values of unimportant model terms. For instance, the SINDy model here exhibits weak effective nonlinearity, $S_e \approx 10^{-5}$, compared with $S_e \approx 10^{-2}$ for the POD-Galerkin model. This small value of $S_e$ may indicate increased difficulty in obtaining a fully negative definite $\bm{A}^S$. SINDy models with weaker sparsity-promotion exhibit larger $S_e$, but then it becomes exceedingly difficult to obtain accurate models in the nonconvex optimization problem. 
Without any sparsity-promotion this is an ill-conditioned, nonconvex optimization in a $330$-dimensional space.  In this way, there appears to be some tradeoff between sparsity-promotion and the strength of effective nonlinearity. 
Given these points, we consider the sparse 5-mode and 9-mode SINDy models to be promising first steps towards incorporating stability constraints into higher-dimensional data-driven models. 

Before concluding, we should note that the eight-mode (no shift mode) POD-Galerkin model from Noack et al.~\cite{noack2003hierarchy}, and all eight-mode models found by  trapping SINDy, do not exhibit global stability. 
The problem fundamentally stems from the marginal stability of the mean flow and the very weak effective nonlinearity, both of which are somewhat addressed by the shift mode in the 9-mode model. 
This should be taken as a cautionary warning; success of these algorithms still relies on useful representations that capture the stability information of the underlying dynamics. 
This may require high-resolution data or the alternative dynamic bases mentioned in Section~\ref{Sec:ProjROMS}.

\section{Conclusion\label{sec:conclusion}}
The present work develops physics-constrained system identification by biasing models towards fulfilling global stability criteria, and subsequently produces long-term bounded models with no extra assumptions about the stability properties of equilibrium points and equilibrium trajectories. In order to produce globally stable models, we have implemented a new trapping SINDy algorithm based on the Schlegel-Noack trapping theorem~\cite{Schlegel2015jfm}. 
Biasing models towards stability, and post-fit, proving that identified models are globally stable, will likely become increasingly important for both projection-based and data-driven models of fluids and plasmas. Our approach, which relies on using the energy as a Lyapunov function for an entire class of models with fixed nonlinear structure, is challenging for application to higher-order nonlinearities where generic Lyapunov functions are often unknown. Fortunately, data-driven methods are now increasingly used to discover Lyapunov functions and barrier functions for nonlinear control~\cite{neumann2013neural,khansari2014learning, richards2018lyapunov,kolter2019learning,jin2020neural,takeishi2020learning,boffi2020learning,chang2020neural,massaroli2020stable}. These methods build a heuristic Lyapunov function for a given dataset, rendering the search for a Lyapunov function tractable but possibly at the cost of model generality. 

We demonstrated the effectiveness of this optimization to identify stable models and additionally managed to improve the discovery of models that do not conform to the assumptions of the trapping theorem. 
Our trapping SINDy algorithm resulted in more accurate and stable models for a range of systems, including simple benchmark problems, noisy data from chaotic systems, and DNS from full spatial-temporal PDEs. 
In these examples, we found that our modified SINDy algorithm could effectively discover stable, accurate, and sparse models from significantly corrupted data. 
Even when an explicit stable trapping region was not found, improved stability was observed. Finally, we explored relatively high-dimensional reduced-order models, with $\mathcal{O}(10)$ degrees of freedom, which are typically challenging for unconstrained data-driven algorithms.  

There is considerable future work for biasing machine learning methods to discover models that satisfy existence-style proofs of stability, especially those that require nonconvex optimization; we find that the lack of convexity in the trapping SINDy algorithm leads to deprecating algorithm speed and tractability as the size of the problem increases. There are many fluid flows which have known stable and unstable projection-based and data-driven reduced-order models, and which would benefit from a larger class of models with trapping region guarantees. Future work should apply this methodology to heavily-researched systems such as the fluidic pinball~\cite{deng2020low,raibaudo2020machine} and the lid-cavity flow~\cite{terragni2011local, lorenzi2016pod}.
Other promising future work includes adapting this structure to N-body coupled Stuart-Landau equations for which stability theorems already exist~\cite{panteley2020practical}. 
However, the nonconvexity of this formulation may require adaptation to a deep learning approach for high-dimensional N-body problems that occur in fluids and modern neuronal models.

For all of the examples in this work, we train our trapping SINDy algorithm on a single trajectory, although most data-driven methods can improve performance by processing data from multiple trajectories. Very large data can be effectively addressed with modern approaches, such as manifold Galerkin projection~\cite{loiseau2019pod} and autoencoder~\cite{baldi1989neural,Milano2002jcp,lusch2018deep,champion2019data,lee2020model} methods.  
These approaches may also address the significant Kolmogorov width limitations of linear transformations~\cite{pinkus2012n}, and help ease the nonconvexity of our new optimization problem. 
There are also modern reduced-order modeling techniques, such as ``lift $\&$ learn''~\cite{qian2020lift}, which produce quadratic ROMs regardless of the nonlinear structure of the underlying governing equations. 
Similarly Koopman analysis aims to produce a map from the original state-space, where the dynamics are nonlinear, to a new coordinate system, typically infinite dimensional, where the dynamics become linear~\cite{koopman_hamiltonian_1931,mezic_analysis_2013,klus2018data,lusch2018deep,li_extended_2017,yeung2019learning,Takeishi2017nips,otto2019linearly}. 
It will be interesting to further explore the connections between the trapping theorem and these related approaches.  
For instance, Pan et al.~\cite{pan2020physics} builds stable Koopman models by requiring that the real parts of the eigenvalues of the linear Koopman operator are non-positive, although the relationship between this linear stability and the trapping theorem is unclear. 
In related work, neural-network-based encoders are often used to reverse this mapping; encoders can input quadratically nonlinear fluid flow data and apply nonlinear transformations to find useful reduced-order models beyond what is capable with traditional projection-based methods~\cite{gonzalez2018deep}. 
A natural question that arises is: assuming the original energy-preserving, quadratically nonlinear fluid flow exhibits a trapping region, under what conditions can we conclude that global stability holds in a new coordinate system given by $\bm{b} = \bm{g}(\bm{y})$? 
The transformation could be an encoder, the reverse lifting map~\cite{qian2020lift}, or some other coordinate transform. 
Understanding how the stability properties manifest in the transformed system is a promising future direction for extending this stability theorem for ROMs with alternative dynamic bases.

\section*{Acknowledgements}
This work was supported by the Army Research Office ({ARO W}911{NF}-19-1-0045) and the Air Force Office of Scientific Research (AFOSR {FA}9550-18-1-0200).
JLC acknowledges support from the National Defense Science and Engineering Fellowship. 
SLB would like to acknowledge valuable discussions with Tulga Ersal, Ronan Fablet, Alireza Goshtabi, Nathan Kutz, JC Loiseau, Bernd Noack, and Said Ouala related to SINDy, Galerkin models, and stability constraints. 

\appendix
\section{Von K\`arm\`an DNS and POD-Galerkin details}\label{appendix:vonKarman_DNS}
We simulate the flow past a circular cylinder at Reynolds number $\rm Re=100$ with unsteady incompressible DNS using the open source spectral element solver Nek5000 \cite{Nek5000}.  
The domain consists of $17,432$ seventh order spectral elements ($\sim 850,000$ grid points)
on $ x, y \in (-20, 50) \times (-20, 20) $, refined close to a cylinder of unit diameter centered at the origin.
Diffusive terms are integrated with third order backwards differentiation, while convective terms are advanced with a third order extrapolation.
The 9-mode augmented POD-Galerkin model is computed following Noack et al~\cite{noack2003hierarchy}, using gradients extracted directly from the DNS code.
Mean-subtracted POD modes are computed from a set of 100 equally-spaced snapshots over one period of vortex shedding.
The shift mode is calculated as the difference between an unsteady base flow, obtained with the selective frequency damping algorithm~\cite{Akervik2006pof} and the mean of the snapshots, orthonormalized with respect to the remaining POD modes with a Gram-Schmidt procedure.
The transient DNS is initialized with the unstable steady state perturbed by the leading linear instability mode with energy $10^{-8}$.
We compute the transient POD coefficients by projecting 3000 snapshots sampled at $\Delta t = 0.1$ onto this POD basis.

 \begin{spacing}{.9}
 \small{
 \setlength{\bibsep}{4.pt}
 \bibliographystyle{unsrt}
 \bibliography{attractor}
 }
 \end{spacing}

\end{document}